\documentclass[10pt,journal,compsoc]{IEEEtran}
\usepackage[nocompress]{cite}

\usepackage{mathrsfs}
\usepackage{amsfonts}
\usepackage{graphicx,epsfig,amssymb,amsmath}
\usepackage{color,xcolor}
\definecolor{red}{rgb}{1.00, 0.00, 0.00}  
\usepackage{pifont}
\usepackage{stmaryrd}
\usepackage{setspace}
\usepackage{subfigure}
\usepackage{array}
\usepackage{float}
\usepackage{multirow}
\usepackage{algorithmic,algorithm}
\usepackage{epstopdf}
\usepackage{mathtools}
\usepackage{diagbox}
\usepackage{booktabs}
\usepackage{verbatim}

\newcommand{\bm}[1]{\mbox{\boldmath{$#1$}}}
\newcommand{\tabincell}[2]{\begin{tabular}{@{}#1@{}}#2\end{tabular}}

\usepackage{arydshln}
\usepackage{enumitem}
\usepackage{ragged2e}
\usepackage{hyperref}
\begin{document}
	
\title{Decentralized Inference with Graph Neural Networks in Wireless Communication Systems}
	
\author{Mengyuan~Lee,~ \IEEEmembership{Student Member,~IEEE,}  Guanding~Yu,~\IEEEmembership{Senior Member,~IEEE,} and~Huaiyu~Dai,~\IEEEmembership{Fellow,~IEEE}
	\IEEEcompsocitemizethanks{
	    \IEEEcompsocthanksitem The work of Guanding Yu was supported by the National Key Research and Development Program of China under Grant 2018YFB1802302. The work of Huaiyu Dai was supported in part by the US National Science Foundation under grant CNS-1824518.
		\IEEEcompsocthanksitem M. Lee  is with the College of Information Science and Electronic Engineering, Zhejiang University, Hangzhou 310027, China, and also with the Department of Electrical and Computer Engineering, North Carolina State University,
		Raleigh, NC 27695, USA. e-mail: mengyuan\_lee@zju.edu.cn. 
		\IEEEcompsocthanksitem G. Yu is with the College of Information Science and Electronic Engineering, Zhejiang University, Hangzhou 310027, China. e-mail: yuguanding@zju.edu.cn. (\emph{Corresponding author: Guanding Yu})
		 \IEEEcompsocthanksitem H. Dai is with the Department of Electrical and Computer Engineering, North Carolina State University,
		 Raleigh, NC 27695, USA. e-mail: hdai@ncsu.edu
}}

\IEEEtitleabstractindextext{
\begin{abstract}
\justifying  
Graph neural network (GNN) is an efficient neural network model for graph data and is widely used in different fields, including wireless communications. Different from other neural network models, GNN can be implemented in a decentralized manner during the inference stage with information exchanges among neighbors, making it a potentially powerful tool for decentralized control in wireless communication systems.  The main bottleneck, however, is wireless channel impairments that deteriorate the  prediction robustness of GNN. To overcome this obstacle, we analyze and enhance the robustness of the decentralized GNN during the inference stage in different wireless communication systems in this paper. Specifically, using a GNN binary classifier as an example,  we first develop a methodology to verify whether the predictions are robust. Then, we analyze the performance of the decentralized GNN binary classifier in both uncoded and coded wireless communication systems. To remedy imperfect wireless transmission and enhance the prediction robustness, we further propose novel retransmission mechanisms for the above two communication systems, respectively. Through simulations on the synthetic graph data, we validate our analysis, verify the effectiveness of the proposed retransmission mechanisms, and provide some insights for practical implementation.
\end{abstract}

\begin{IEEEkeywords}
Machine learning, graph neural network, decentralized algorithm, retransmission, robustness.
\end{IEEEkeywords}}

\maketitle

\IEEEraisesectionheading{\section{Introduction}}
\IEEEPARstart{W}{ith} the explosion of data traffic and the enhanced capability of big data processing, the machine learning (ML) technique has been regarded as a potential enabler for the sixth-generation (6G) wireless communications \cite{6g}. Recently, a large number of works turn to various ML techniques to facilitate different applications in wireless communication systems, such as 
physical layer design \cite{physical1,physical2,physical3,deepfolding}, resource allocation\cite{shi,lorm, learntobranch,spatial,graphnn,mlop}, and networking \cite{networking1,networking2}. In these works, ML is generally utilized for two purposes. On the one hand, it has been employed as fast approximations of heavy computations in existing algorithms. For example, the authors in \cite{lorm,learntobranch} have made use of the imitation learning technique to replace the pruning strategy in the branch-and-bound algorithm, a widely-used algorithm for mixed-integer wireless  resource allocation problems with exponential complexity in the worst case. On the other hand, ML has also been used to develop novel low-complexity heuristic algorithms. For example, the authors in \cite{spatial} and \cite{graphnn} have proposed ML-based link scheduling algorithms for device-to-device networks by using deep neural networks (DNNs) and the graph embedding technique, respectively.

Amongst other ML techniques, the graph neural network (GNN) recently attracts much attention from both the academia and the industry. As the name implies, GNN is a kind of neural network model especially proposed for graph data \cite{gnn_survey1,gnn_survey2}. It can learn the representation of an input graph in the node/graph level via message passing between nodes. Because of its satisfactory performance and good interpretability, GNN has become a widely applied graph analytical tool in various fields, such as computer vision \cite{cv}, traffic engineering \cite{traffic}, and chemical industry \cite{chem}. By observations, GNN has several properties that highly match with the characteristics/requirements of the applications in wireless communication systems. First, GNN can efficiently exploit the graph data for different learning tasks while wireless networks can be naturally modeled as graphs, where nodes and edges correspond to wireless devices and links, respectively. Secondly, GNN can process input graphs of different sizes.  This generalization ability is in accord with the dynamic nature of wireless communication systems. Finally, after a well-trained GNN model is obtained\footnote{The training process of the GNN can be conducted in either centralized or decentralized manner by using different training techniques. Since using a well-trained GNN model, i.e., the inference stage of the GNN, is the focus of this paper, we assume that a well-trained GNN model is available throughout this paper.}, it can be used in a decentralized manner for inference to achieve decentralized management/control.  Meanwhile, decentralized control and resource management often serves as an appealing alternative especially for large-scale wireless communication systems  \cite{de1,de2,de3}, which does not depend on the centralized controller, and thus avoiding concurrent communication, possibly severe traffic jam, the computation limitation, low scalability and flexibility, and single point of failure. Note that most standard neural networks widely used in existing works, such as convolutional neural networks (CNNs) and recurrent neural networks (RNNs), requires central processing during the inference stage.

To employ the inference stage of a well-trained GNN model decentrally in wireless communication systems, information exchanges among neighboring nodes through wireless channels are inevitable. However, wireless transmission over noisy and fading channels is generally imperfect, which leads to stochastic errors in received signals, and eventually decreases the prediction accuracy and deteriorates the robustness of the GNN model. This is the main bottleneck concerning decentralized inference with GNN in wireless systems. Therefore, the goal of this paper is to analyze and enhance the robustness of the decentralized GNN during the inference stage in different wireless communication systems, making the prediction results not only accurate but also robust to transmission errors.

\subsection{Related Work}
\subsubsection{Related Work in Machine Learning Field}
GNNs, as the generalization of traditional DNNs to the graph data, inherit both advantages and disadvantages of DNNs. Specifically, GNNs have powerful representation ability, but are vulnerable to adversarial attacks during the inference stage\footnote{Adversarial attacks can also happen during the training stage of GNNs, which is out of the scope of this paper.}. Note that adversarial attacks refer to the manually designed perturbations on the node/edge features or the topology of the graph data, which aim at fooling GNNs to get incorrect predictions. Therefore, there exist a surge of works in the ML field about enhancing the robustness of GNNs against adversarial attacks during the inference stage \cite{attack,adver_training1,adver_training2,asdver_detec1,asdver_detec2,model_improv1,model_improv2}. Existing adversarial defense methods can be categorized into three types \cite{attack}: (i) adversarial training that injects adversarial samples into the training set such that the GNN model can correctly predict the label of testing samples with adversarial attacks \cite{adver_training1,adver_training2}, (ii) adversary detection that explores the intrinsic differences of adversarial edges/nodes and then cleans edges/nodes to get correct predictions \cite{asdver_detec1,asdver_detec2}, and (iii) model improvement that focuses on improving the GNN model itself to make it robust against the adversarial attacks \cite{model_improv1,model_improv2}. There also exist some works that theoretically analyze the robustness of GNNs under adversarial attacks \cite{robust_1,robust_2}. However, existing adversarial attacks in literature are manually designed based on certain criteria, which are quite different from the random and stochastic errors caused by imperfect wireless transmission. Therefore, all aforementioned works about defending adversarial attacks may not be directly applied to the decentralized GNN in wireless communication systems. This motivates us to study and enhance the robustness of the decentralized GNN in wireless communication systems.

\subsubsection{Related Work in Wireless Communication Field} 
In the wireless communication field, related works can be divided into two categories. The first category of works focus on using GNNs for different problems in wireless communication systems, such as link scheduling \cite{graphnn}, power allocation \cite{eisen_graph,shen_graph} and user association \cite{liu_graph}. In these works, GNNs are generally used in the centralized manner. Recently, some works have used GNNs for decentralized control \cite{control1,control2,control3}. However, they generally assume perfect information exchanges between nodes. In contrast, the second category of works examine how wireless transmission affects ML performances \cite{fl_survey,chen_fl,vin_fl,ren,adptive_fl,huang_retrans}. To be more specific, they focus on the training process of ML models and analyze the convergence performance of training algorithms, such as the stochastic gradient descent (SGD) algorithm, over noisy wireless channels. Based on the analysis, resource management algorithms \cite{chen_fl,vin_fl}, learning parameter setting methods \cite{ren,adptive_fl}, as well as scheduling mechanisms \cite{huang_retrans} are proposed to achieve different goals, such as trade-off between communication and computation resources, faster convergence, or better ML performance. These existing methods are also applicable to the decentralized training of the GNN in wireless communication systems. However, the characteristics of the training and inference stages of ML are very different. Specifically, training an ML model using algorithms, such as SGD, is a stochastic process, which is inherently tolerant of noisy data. On the contrary, using an ML model for prediction is a definite process and is more vulnerable to noise. Therefore, these aforementioned works are not applicable in our considered problem, which motivates the study in this paper.

\subsection{Contributions and Outline}
\noindent As far as we know, we are among the first to study robust decentralized inference with GNNs in wireless communication systems. To summarize, our main contributions are as follows.
\begin{itemize}[leftmargin=3mm]
	\item We formulate  and solve a robustness verification problem to check whether the predictions of the decentralized GNN binary classifier in wireless communication systems are robust.
	\item We analyze the performance of the decentralized GNN binary classifier in both uncoded and coded communication systems by examining the robustness verification problem.
	\item We develop two novel retransmission mechanisms based on the combining technique to deal with the impairments of wireless transmission and enhance the robustness of the decentralized GNN binary classifier in uncoded and coded communication systems, respectively.
	\item We conduct extensive simulations through which we reveal the effectiveness of our proposed retransmission mechanisms and shed lights on practical implementation.
\end{itemize}

The rest of this paper is organized as follows. In Section \ref{s2}, we give a brief overview of GNNs. In Section \ref{s3}, we introduce the binary classification problem in wireless communication systems and discuss a corresponding GNN binary classifier as well as its decentralized implementation. In Section \ref{s5}, we formulate and solve the robustness verification problem. Based on this, we analyze  and enhance the robustness of the decentralized GNN binary classifier in uncoded and coded communication systems in Sections \ref{s6} and \ref{s7}, respectively. Extensive testing results are presented in Section \ref{s8}. Finally, we conclude the paper and provide directions for future work in Section \ref{s9}.

\subsection{Notations} 
\noindent Throughout this paper, vectors are denoted in bold lower-case letters, such as $\bm{a}$, while matrices are denoted in bold upper-case letters, such as $\bm{A}$. Moreover, we use $\bm{a}[i]$ to denote the $i$-th entry in vector $\bm{a}$ and $\bm{A}[i,j]$ to denote the entry in the $i$-th row and $j$-th column of matrix $\bm{A}$. We also use $\bm{A}[i,:]$ and $\bm{A}[:,j]$ to denote the $i$-th row and the $j$-th column in $\bm{A}$, respectively. Given two sets, $S_1$ and $S_2$, $\bm{A}_{\{S_1,S_2\}}$ corresponds to a sliced matrix from $\bm{A}$, which contains the rows corresponding to the indices  in $S_1$ and the columns corresponding to the indices  in $S_2$. Furthermore, given two matrices $\bm{A}$ and $\bm{B}$, if they have the same number of rows, we can stack them in column and the column-stacking result is denoted as $[A,B]$. Similarly, if they have the same number of columns, we can stack them in row and the row-stacking result is denoted as $[A;B]$. For example, assuming that the dimensions of $\bm{A}$ and $\bm{B}$ are both $2\times 2$, then we have
\begin{small}
	\begin{equation*}
	[A,B]
	=
	\left[
	\begin{array}{cccc}
	A[1,1], & A[1,2], & B[1,1], & B[1,2] \\
	A[2,1], & A[2,2], & B[2,1],& B[2,2]  \\
	\end{array}
	\right],
	\end{equation*}
	\begin{equation*}
	[A;B]
	=
	\left[
	\begin{array}{cc}
	A[1,1], & A[1,2]  \\
	A[2,1], & A[2,2] \\
	B[1,1], & B[1,2]\\
	B[2,1],& B[2,2] 
	\end{array}
	\right].
	\end{equation*}
\end{small}
\vspace{-2em}
\section{Overview of Graph Neural Networks} \label{s2}
\noindent GNNs, to some extent, are inspired by the success of CNNs that reduce the number of parameters in fully-connected DNNs with convolution filters. However, the convolution operation in CNNs only works for grid-like image data over the Euclidean space. Therefore, GNNs are proposed by redefining the convolution operation for graphs \cite{gnn_survey1,gnn_survey2}. To discuss the typical architecture of GNNs, we consider a graph $G(\mathcal{V},\mathcal{E})$ with the node set $\mathcal{V}$ and the edge set $\mathcal{E}$. The goal of GNNs is to learn a state embedding vector $\bm{h}_v\in \mathbb{R}^D$ for each node $v\in\mathcal{V}$ by taking the graph adjacency matrix, node features, and edge features (if available) as inputs. Note that $D$ is a hyperparameter called hidden state dimension whose value is manually set. Then the state embedding vector $\bm{h}_v$ is used to generate the output of node $v$, i.e., $\bm{o}_v$. Generally, a GNN is constructed with a sequence of $K$ layers. In the $k$-th layer, the hidden state of node $v$, denoted as $\bm{h}_v^{(k)}$, is first updated by collecting the neighborhood information as follows 
\begin{equation}
\bm{h}_v^{(k)} = f_t^{(k)}(\bm{x}_v,\bm{x}_{co[v]},\bm{x}_{ne[v]}, \bm{h}^{(k-1)}_{ne[v]}; \bm{\theta}_t^{(k)}), \label{local_tran}
\end{equation}
where $f_t^{(k)}$ is called the local transition function in the $k$-th layer. Note that $f_t^{(k)}$ is the same for all nodes in $\mathcal{V}$ and generally realized through a multi-layer perception (MLP), whose learning parameters are denoted as $\bm{\theta}_t^{(k)}$. Meanwhile, $\bm{x}_v$, $\bm{x}_{co[v]}$, $\bm{x}_{ne[v]}$ are the features of node $v$, its edges, and its neighbors, respectively.  Also, $\bm{h}^{(k-1)}_{ne[v]}$ denotes the hidden states of the neighbors of node $v$ in the $(k-1)$-th layer. After the update in (\ref{local_tran}), the output of node $v$ in the $k$-th layer, denoted as $\bm{o}_v^{(k)}$, is computed as
\begin{equation}
\bm{o}_v^{(k)} = f_o^{(k)}(\bm{x}_v,\bm{h}^{(k)}_{v};\bm{\theta}_o^{(k)}),\label{local_out} 
\end{equation}
where $f_o^{(k)}$ is called the local output function in the $k$-th layer, which defines how the output of each node is produced. Similar to $f_t^{(k)}$, $f_o^{(k)}$ is again the same for all nodes in $\mathcal{V}$ and generally realized through an MLP with learning parameters, $\bm{\theta}_o^{(k)}$ \footnote{Eq.  (\ref{local_tran}) and (\ref{local_out}) are general formulas of the local transition and output functions. The specific design of these two functions can be different in various GNN models by omitting some input components, adding extra components, choosing different techniques to combine input components, or selecting different MLP structures.}.

As mentioned above, GNN has many special properties that match the characteristics and requirements of wireless communication systems. One important property is that  the inference stage of the GNN can be naturally implemented in a decentralized manner. Specifically, the operations in (\ref{local_tran}) and (\ref{local_out}) only need the information of node $v$ ($\bm{x}_v$, $\bm{x}_{co[v]}$, and $\bm{h}_v^{(k)}$) and its one-hop neighbors ($\bm{x}_{ne[v]}$ and $\bm{h}_{ne[v]}^{(k-1)}$), where the one-hop neighbors' information can be obtained by local information exchanges. In other words, operations in each layer of the GNN are entirely local. However, in wireless communication systems, the information exchanges among neighbors are achieved through noisy and fading channels, whose impairments lead to imperfect transmission and deteriorate the robustness of decentralized predicted results. This is the main bottleneck for the decentralized inference with the GNN.  Analyzing the performance of the decentralized GNN in different wireless communication systems and overcoming the above obstacles to achieve robust predictions are the focus of this paper.

\section{System Model and GNN Binary Classifier} \label{s3}
\noindent For simplicity, we focus on a decentralized GNN binary classifier in the subsequent discussion of this paper\footnote{The proposed analysis and retransmission mechanism can be generalized to more complicated problems by modifying the GNN structure and the formulation of the robustness verification problem accordingly.}. In this section, we first introduce the binary classification problem in wireless communication systems. Then, we discuss the architecture of a GNN binary classifier for this problem as well as its decentralized implementation in wireless communication systems.

\subsection{Binary Classification Problem} \label{s3_1}
\begin{figure}
	\vspace{-1em}
	\centering
	\includegraphics[width=0.8\linewidth, height=0.15\textheight]{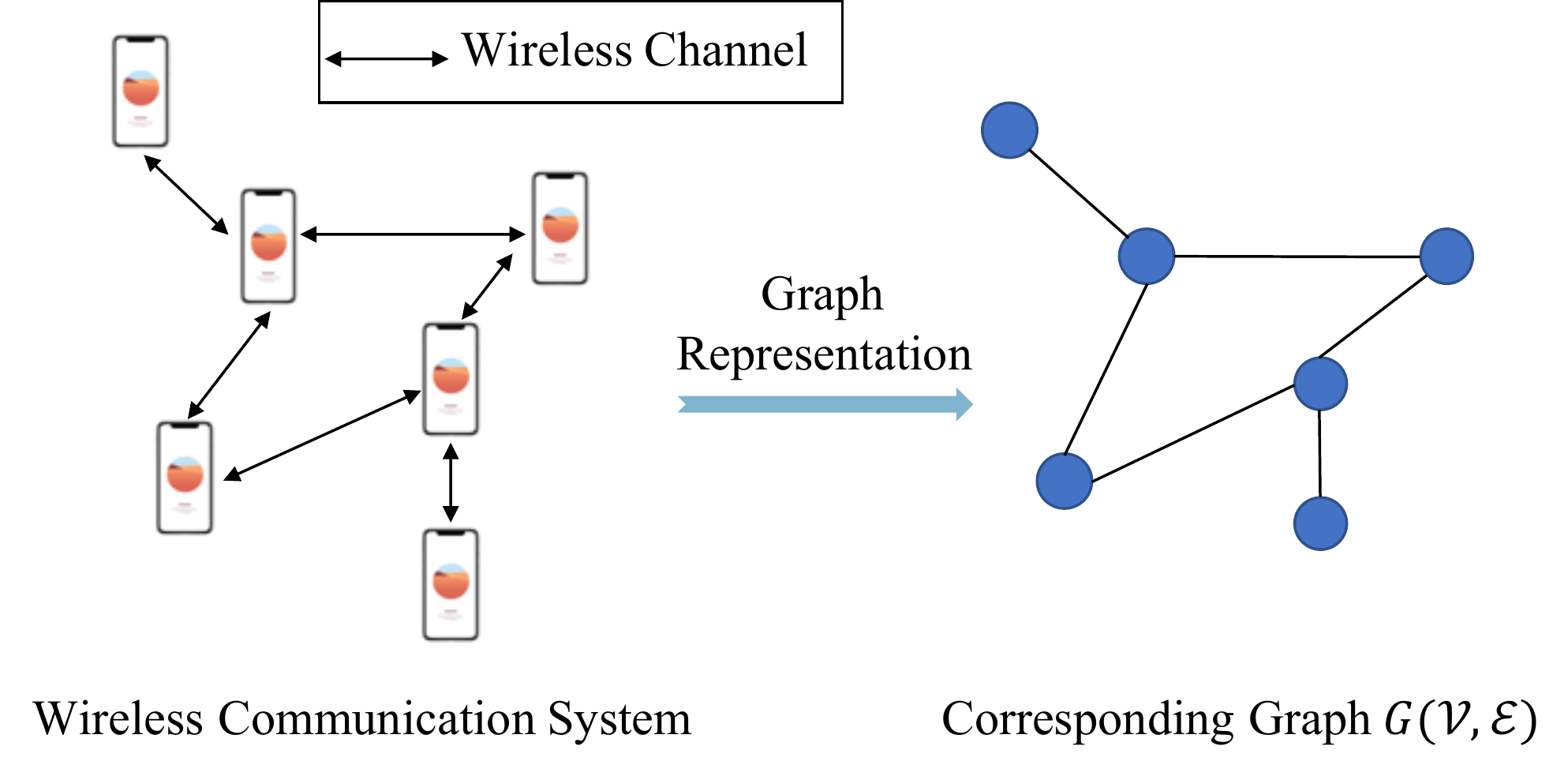}
	\vspace{-1em}
	\caption{A general wireless communication system with $N=6$ nodes and its corresponding graph model.}
	\label{fig:system}
	\vspace{-2em}
\end{figure}

\noindent As shown in Fig. \ref{fig:system}, we consider a binary classification problem in the wireless communication system that can be widely found in literature \cite{shi,spatial,graphnn,de3} and  modeled as a graph $G(\mathcal{V},\mathcal{E})$, with the node set $\mathcal{V}$ and the edge set $\mathcal{E}$. Specifically, this system has $N$ nodes. These nodes can be base stations, access points, edge servers, or mobile devices according to specific applications. To enable local computation, we assume that each device is  equipped with sufficient storage and computing capability. Meanwhile, two nodes are neighbors and connected by an edge if they can communicate with each other through the wireless channel.  In this paper, we assume that each node knows its neighbors in the system, which can be achieved by neighbor discovery algorithms such as \cite{neighbor_discovery}. Moreover, each node in $\mathcal{V}$ can be classified into two classes and correspondingly labeled as 1 or -1. The label refers to binary indicators in wireless communication systems, such as the power allocation indicator in the power control problem  \cite{shi}, the link activation indicator in the wireless link scheduling problem \cite{spatial}, and the subcarrier assignment indicator for the subcarrier assignment problem \cite{de3}. Furthermore, each node $v \in \mathcal{V}$ keeps a vector that includes the needed internal characteristics for the classification task, such as the transmit power limit or the QoS requirement for the power control problem \cite{shi}. For the sake of storage and digital communication, this vector is quantized into a binary node feature vector, $\bm{x}_v \in \{0,1\}^p$, where $p$ is called the node feature dimension and is manually set based on the requirements of the quantization precision. In this paper, we do not take the detailed quantization process into consideration and assume that $\bm{x}_v$ is already available at each node.

\vspace{-1em}
\subsection{Wireless Communication Model} \label{s3_wireless}
\noindent Assume that node $v$ and node $u$ are neighbors. Then, when node $u$ transmits a $p$-bit signal $\bm{x}_u$ to node $v$,  the transmission takes $p$ time slots that is defined as a symbol block. Furthermore, we assume that the transmission of $\bm{x}_u$ is not only  contaminated by the additive white Gaussian noise (AWGN) that is independent and identically distributed (i.i.d.) over different time slots, but also impaired by the pathloss, the shadowing, and the block-Rayleigh fading, which indicates that the channel gain remains static within a symbol block but is i.i.d over different blocks. According to \cite{huang_retrans}, we further assume that the receiver $v$ can estimate the signal-to-interference-plus-noise ratio  (SINR)  $\gamma_{vu}$ of the received signal from each neighbor $u \in N(v)$.

Two kinds of communication systems, uncoded and coded communication systems, are considered in this paper. If the transmitted signal from node $u$, i.e., $\bm{x}_u$, is  uncoded,  node $v$ can use $\gamma_{vu}$ to estimate the bit error rate (BER), $\epsilon_{vu}$, of the received signal. On the other hand, if $\bm{x}_u$ is coded before transmission, we assume that it can be correctly decoded by node $v$  if $\log(1+\gamma_{vu}) \geq R_u$, where $R_u$ is the preset transmit data rate of node $u$. Otherwise, the received signal is said to be in outage and the whole packet will be discarded.

\subsection{Architecture of GNN Binary Classifier}\label{s4_1}
The aforementioned binary classification problem can be solved by a GNN binary classifier. The operations of the GNN binary classifier are as follows.  First, the hidden states of all nodes are updated  as 
\begin{equation}
\bm{H} = \sigma^{(r)}(\bm{\hat{A}}\bm{X}\bm{\theta}) \triangleq F(\bm{\hat{A}},\bm{X};\bm{\theta}), \label{Cgnn} 
\end{equation}
where $\bm{X}\in \{0,1\}^{N\times p}$ is the node feature matrix, which is obtained by row-stacking node features of all nodes in $\mathcal{V}$. Also, $\bm{\theta} \in \mathbb{R}^{p \times D}$ is the matrix of learning parameters, $\sigma^{(r)}(\cdot)$ is the ReLU function, and $\bm{H} \in \mathbb{R}^{N\times D}$ is the embedding output matrix that is obtained by row-stacking the embedding output vectors of all nodes, i.e., $\{\bm{h}_v, \forall v \in \mathcal{V}\}$. Furthermore, $\bm{\hat{A}} \in \mathbb{R}^{N\times N}$ is called graph filter, which defines how to integrate neighborhood information. It is generally a variant of the graph adjacency matrix $\bm{A} \in \{0,1\}^{N\times N}$. There are three popular graph filters for GNNs as follows
\begin{itemize}
	\item \textbf{Unnormalized graph filters} \cite{unnormal}: $\bm{\hat{A}} = \bm{A}+\bm{I}_N$, where $\bm{I}_N$ is the identity matrix.
	\item \textbf{Normalized graph filters} \cite{gcn}: $\bm{\hat{A}} = \bm{D}^{-1/2}\bm{A}\bm{D}^{-1/2}+\bm{I}_N$, where $\bm{D}$ is the degree matrix. 
	\item  \textbf{Random walk graph filters} \cite{random_walk}: $\bm{\hat{A}} = \bm{D}^{-1}\bm{A}+\bm{I}_N$.
\end{itemize}
As mentioned in Section \ref{s3_1}, the neighborhood information is available for each node. Therefore, the needed graph filter information can be assumed to be available for each node as well.

After the operation in (\ref{Cgnn}), the hidden states of all nodes are input into a fully-connected layer for binary classification. The specific operation is given by 
\begin{equation}
\bm{y}=\sigma^{(s)}(\bm{H}\bm{w}+b) \triangleq G(\bm{H};\bm{w},b), \label{Cmlp} 
\end{equation}
where $\bm{y} \in [0,1]^N$ is the predicted output vector of all nodes, with each entry indicating the probability that the corresponding node is labeled as 1. Also, $\sigma^{(s)}(\cdot)$ is the sigmoid function, $\bm{w} \in \mathbb{R}^D$ is the vector of learning weights, and $b$ is the bias. We use $\mathcal{W}= {\{\bm{\theta},\bm{w},b\}}$ to denote the set of all learning parameters in the GNN binary classifier. 

Based on the predicted output $\bm{y}$, we can get the predicted binary label vector of all nodes that is denoted as $\bm{c} \in \{-1,1\}^N$ as 
\begin{equation}
\bm{c} = sgn (\bm{y}-0.5). \label{Clabel}
\end{equation}
By integrating the operations in (\ref{Cgnn}), (\ref{Cmlp}), and (\ref{Clabel}), operations of the GNN binary classifier can be overall denoted as 
\begin{equation}
\bm{c}=J(\bm{\hat{A}},\bm{X};\mathcal{W}). \label{C_all}
\end{equation}

\subsection{Decentralized Implementation of GNN Binary Classifier}  \label{s4_2}
As mentioned above, decentralized management/control is preferred in wireless communication systems. Therefore, in this part, we discuss the decentralized implementation of the above GNN binary classifier in the wireless communication system. As a premise, we assume that the GNN binary classifier is already well-trained and each node $v \in \mathcal{V}$ keeps a copy of it, i.e. $J(\cdot, \cdot;\mathcal{W})$, which indicates that the operations and parameters of the GNN model are available at each node. In the remaining part of this subsection, we introduce how each node uses $J(\cdot, \cdot;\mathcal{W})$ to predict its label in a decentralized manner. 

Note that the operations (\ref{Cgnn})-(\ref{Clabel}) of the GNN binary classifier introduced in the previous subsection are given in the centralized manner, where the graph filter $\bm{\hat{A}}$ and the node feature matrix $\bm{X}$ of all the nodes are needed. To better illustrate the decentralized implementation of this GNN binary classifier, we need to manifest  the operations in (\ref{Cgnn})-(\ref{Clabel}) for each single node $v \in \mathcal{V}$. First, we denote the set of one-hop neighbors of node $v$ as $N(v)$. We further define the sliced graph filter, $\bm{\hat{a}}_v \in \mathbb{R}^{1\times|N(v)|}$, as $\bm{\hat{a}}_v=\bm{\hat{A}}[v,:]_{\{N(v)\}}$, which keeps the graph filter entries needed by node $v$ to integrate information from its neighbors. Similarly, we define the sliced node feature matrix, $\bm{X}_v \in \{0,1\}^{|N(v)|\times p}$, as $\bm{X}_v=\bm{X}_{\{N(v),:\}}$, which contains the node features of nodes in $N(v)$, i.e., $\{\bm{x}_u, \forall u \in N(v)\}$. With these newly defined notations, the operations in  (\ref{Cgnn})-(\ref{Clabel}) can be rewritten for node $v$ as
\begin{equation}
\begin{aligned}
\bm{h}_v &=F([\bm{\hat{a}}_v,\hat{a}_v],[\bm{X}_v;\bm{x}_v];\bm{\theta}) \\&=\sigma^{(r)}([\bm{\hat{a}}_v,\hat{a}_v][\bm{X}_v;\bm{x}_v]\bm{\theta})\\&= \sigma^{(r)}(\hat{a}_v\bm{x}_v\bm{\theta}+\bm{\hat{a}}_v\bm{X}_v\bm{\theta}),\label{Lgnn} 
\end{aligned}
\end{equation}
\begin{equation}
y_v= G(\bm{h}_v;\bm{w},b)=\sigma^{(s)}(\bm{h}_v\bm{w}+b), \label{Lmlp} 
\end{equation}
\begin{equation}
c_v = sgn (y_v-0.5), \label{Llabel}
\end{equation}
where $\hat{a}_v=\bm{\hat{A}}[v,v]$, $\bm{h}_v \in \mathbb{R}^{1\times D}$ is the embedding output vector of node $v$, $y_v \in[0,1]$ is the predicted output value of node $v$, and $c_v \in \{-1,1\}$ is the binary predicted label of node $v$. Thus, the overall operations at node $v$ is given by
\begin{equation}
c_v=J([\bm{\hat{a}}_v,\hat{a}_v],[\bm{X}_v;\bm{x}_v];\mathcal{W}).  \label{L_all}
\end{equation}

From (\ref{L_all}), the graph filter information, the node feature $\bm{x}_v$, and the neighbors' node features $\bm{X}_v=\{\bm{x}_u, \forall u \in N(v)\}$ are needed for node $v$ to get its predicted label. As mentioned above, the graph filter information and $\bm{x}_v$ are already available at node $v$. Given that $\{\bm{x}_u, \forall u \in N(v)\}$ can be obtained by information exchanges with its neighbors, the central unit is not needed and node $v$ can obtain its predicted label in a decentralized manner. 

The specific process of the decentralized prediction for each node $v \in \mathcal{V}$ is as follows. First, node $v$ sends transmission requests to each node $u$ in $N(v)$. After receiving the request from node $v$, each node $u$ in $N(v)$ transmits its node feature $\bm{x}_u$ to node $v$ through wireless channels as shown in Fig. \ref{fig:system}. Given that $\bm{x}_u$ is binary and has a length of $p$, we utilize the binary phase shift keying (BPSK) modulation and the transmission of $\bm{x}_u$  takes $p$ time slots or a symbol block. We denote the signal received by node $v$ from node $u$ as $\bm{\hat{x}}_{vu}$, whose SINR is denoted as $\gamma_{vu}$. Due to imperfect wireless transmission, $\bm{\hat{x}}_{vu}$ is generally not equal to $\bm{x}_u$. By row stacking  $\bm{\hat{x}}_{vu}$ for all nodes in $N(v)$, node $v$ gets an estimated  sliced node feature matrix, $\bm{\hat{X}}_v$. Then node $v$ inputs the graph filter information, $\bm{x}_v$, and $\bm{\hat{X}}_v$ into its local GNN copy $J(\cdot,\cdot;\mathcal{W})$. By the operations in (\ref{L_all}), the decentrally predicted label of node $v$ is finally obtained as $\hat{c}_v =J([\bm{\hat{a}}_v,\hat{a}_v],[\bm{\hat{X}}_v;\bm{x}_v];\mathcal{W})$. 

\section{Robustness Verification Problem} \label{s5}
\noindent As mentioned above, the received sliced node feature matrix, $\bm{\hat{X}}_v$, is generally not equal to the true node feature matrix, $\bm{X}_v$, due to imperfect wireless transmission. It indicates a possibility that the predicted label of node $v$, $\hat{c}_v$, is not robust. Therefore, we need to develop a methodology to verify whether a predicted label is robust before studying the performance of the decentralized GNN binary classifier in wireless communication systems. To this end, we investigate the robustness verification problem in this section.

\subsection{Problem Formulation} \label{s5_1}
\noindent Given that $\bm{X}_v$ and $c_v$ are not available during the inference stage, we cannot directly compare $\hat{c}_v$ with $c_v$ for robustness verification. However, for each received signal $\bm{\hat{x}}_{vu}$, its SINR is available at node $v$, based on which the possible number of errors in $\bm{\hat{x}}_{vu}$ can be estimated. Therefore, we introduce a new vector, $\bm{q}_v=[q_{vu}, \forall u \in N(v)]$, as the local error bound vector of node $v$, where each entry $q_{vu}$ represents the maximum number of tolerable errors in $\bm{\hat{x}}_{vu}$. Obviously, the value of $\bm{q}_v$ is highly related to the specific communication model. In the remaining part of this section, we assume that the value of $\bm{q}_v$ is available and more details about estimating the value of $\bm{q}_v$ will be discussed in Section \ref{s6}. Under this assumption, it is obvious that the true sliced node feature matrix, $\bm{X}_v$, must satisfy the following constraints
\begin{equation} \label{cons_x}
\left\{
\begin{array}{lr}
||\bm{X}_v[u,:]-\bm{\hat{X}}_v[u,:]||_0 \leq q_{vu}, \forall u\in N(v), &\\
\bm{X}_v[u,j] \in \{0,1\}, \forall u\in N(v),  j \in \{1, 2, .., p\},&  
\end{array}
\right. 
\end{equation}
where $||\cdot||_0$ means the L0 norm of the input vector. Therefore, we can search all possible matrices satisfying constraints in (\ref{cons_x}), and then pass them through the GNN model. If the predicted labels with all possible matrices are always the same as $\hat{c}_v$, then the true predicted label $c_v$ must be equal to $\hat{c}_v$. In this sense, $\hat{c}_v$ is considered robust.

Accordingly, the robustness verification problem\footnote{The proposed ``robustness verification problem" does not belong to the robust optimization field, where the objective and constraints are uncertain \cite{robust_opt}.} can be formulated as follows 
\begin{equation} \label{Mrobust}
z(\bm{q}_v,\bm{\hat{X}}_v) \triangleq \min_{\bm{\tilde{X}}_v } \hat{c}_v J([\bm{\hat{a}}_v,\hat{a}_v],[\bm{\tilde{X}}_v;\bm{x}_v];\mathcal{W}), \vspace{-0.5em}
\end{equation}
subject to
\begin{align}
||\bm{\tilde{X}}_v[u,:]-\bm{\hat{X}}_v[u,:]||_0 \leq q_{vu}, \forall u\in N(v), \label{robust_1}\tag{\theequation a} \\
\bm{\tilde{X}}_v[u,j] \in \{0,1\}, \forall u\in N(v),  j \in \{1, 2, .., p\}, \label{robust_2}\tag{\theequation b}
\end{align}
where $\bm{\tilde{X}}_v$ is the optimization variable. If $z(\bm{q}_v,\bm{\hat{X}}_v)>0$, the predicted label $\hat{c}_v$ is robust.

\noindent \emph{Remark 1}: If $z(\bm{q}_v,\bm{\hat{X}}_v)>0$,  $J([\bm{\hat{a}}_v, \hat{a}_v], [\bm{\tilde{X}}_v; \bm{x}_v]; \mathcal{W})$ always has the same sign as $\hat{c}_v$ and is also equal to $\hat{c}_v$ due to the fact that the predicted label can only be 1 or -1.

\subsection{Problem Transformation}  \label{s5_2}
\noindent Obviously, Problem (\ref{Mrobust}) is intractable for two main reasons. On the one hand, it is a combinatorial optimization problem due to the constraint (\ref{robust_2}). On the other hand, even if we relax the binary variables into continuous ones in $[0,1]$, the optimization problem is still non-convex due to non-linear activation functions in the GNN binary classifier, i.e., the ReLU function and the sigmoid function. In this subsection, we transform Problem (\ref{Mrobust}) into a more tractable form with different approximation and relaxation techniques.

\subsubsection{Transformation on Sigmoid Function}  \label{s5_2_1}
According to the definition of the sigmoid function and the binary label transformation rule, the predicted label $c_v$ always has the same sign as $(\bm{h}_v\bm{w}+b)$. Therefore, Problem (\ref{Mrobust}) can be simplified as \vspace{-0.5em}
\begin{equation}\label{Mrobust1}
\hat{z}(\bm{q}_v,\bm{\hat{X}}_v) \triangleq \min_{\{\bm{\tilde{X}}_v, \bm{h}_v \}}  \hat{c}_v(\bm{h}_v\bm{w}+b),
\end{equation}
subject to (\ref{robust_1}), (\ref{robust_2}), and
\begin{align}
\bm{h}_v = \sigma^{(r)}(\hat{a}_v\bm{x}_v\bm{\theta}+\bm{\hat{a}}_v\bm{\tilde{X}}_v\bm{\theta}), \label{robust1_2} \tag{\theequation a}
\end{align}
where the non-linear sigmoid function is removed. Obviously, $\hat{z}(\bm{q}_v,\bm{\hat{X}}_v)$ always has the same sign as $z(\bm{q}_v,\bm{\hat{X}}_v)$. Therefore, by solving problem  (\ref{Mrobust1}), we can also verify that the predicted label $\hat{c}_v$ is robust if $\hat{z}(\bm{q}_v,\bm{\hat{X}}_v)>0$.

\subsubsection{Approximation on ReLU Function}  \label{s5_2_2}
To deal with the non-linearity brought by the ReLU function, we adopt the linear approximation of the ReLU function proposed in \cite{relu_outer}. 

First, we denote the input to the ReLU function in  (\ref{robust1_2}) as $\bm{\hat{h}}_v$ and rewrite (\ref{robust1_2}) into \vspace{-0.5em}
\begin{equation} 
\bm{\hat{h}}_v=\hat{a}_v\bm{x}_v\bm{\theta}+\bm{\hat{a}}_v\bm{\tilde{X}}_v\bm{\theta},\label{relu_h}
\end{equation}
\begin{equation} 
\bm{h}_v = \sigma^{(r)}(\bm{\hat{h}}_v). \label{relu_h_hat} \vspace{-0.5em}
\end{equation}
Then we try to bound the value of each entry $\bm{\hat{h}_v}[i]$ in vector $\bm{\hat{h}_v}$. Specifically, for $\bm{\hat{h}_v}[i]$, we can find its upper bound, $\bm{u}_v[i]$, according to \cite{robust_1} as
\vspace{-0.5em}
\begin{equation}
\bm{u}_v[i] = (\hat{a}_v\bm{x}_v\bm{\theta})[i]+(\bm{\hat{a}}_v\bm{\hat{X}}_v\bm{\theta})[i]+\sum_{u\in N(v)}\sum_{k=1}^{q_{vu}} \bm{\hat{a}}_v[u]\bm{\hat{u}}_v[u,i,k], \label{upper1}
\end{equation}
\begin{equation}
\begin{aligned}
\bm{\hat{u}}_v[u,i,k]=\text{$k$-th\_largest} \{(1-\bm{\hat{X}}_v[u,:])\odot[\bm{\theta}[:,i]]_+\\+\bm{\hat{X}}_v[u,:]\odot[\bm{\theta}[:,i]]_-\},\label{upper2}
\end{aligned}
\end{equation}
where $\text{$k$-th\_largest}(\cdot)$ is a function that can find the $k$-th largest entry in the input vector, $\odot$ denotes element-wise product, $[x]_+=\max\{x,0\}$, and $[x]_-=-\min\{x,0\}$. The first two terms in (\ref{upper1}) correspond to the value of $\bm{\hat{h}}_v[i]$ if $\bm{\tilde{X}}_v=\bm{\hat{X}}_v$. Meanwhile, the last term in (\ref{upper1}) denotes the maximal increase on the value of $\bm{\hat{h}}_v[i]$ for any feasible $\bm{\tilde{X}}_v$ constrained by (\ref{robust_1}) and (\ref{robust_2}), which can be explained as follows. First,  the value of $\bm{\hat{h}}_v[i]$ depends on the aggregation over all neighbors, which accounts for the  summation over $u \in N(v)$. Moreover, for each neighbor $u$ and each bit $\bm{\hat{X}}_v[u,j]$ in $\bm{\hat{X}}_v[u,:]$, if (i) $\bm{\hat{X}}_v[u,j]=0$ and $\bm{\theta}[j,i]>0$, or (ii) $\bm{\hat{X}}_v[u,j]=1$ and $\bm{\theta}[j,i]<0$, flipping $\bm{\hat{X}}_v[u,j]$ brings an increase on $\bm{\hat{h}}_v[i]$. These two conditions can be overall written as  $(1-\bm{\hat{X}}_v[u,:])\odot[\bm{\theta}[:,i]]_++\bm{\hat{X}}_v[u,:]\odot[\bm{\theta}[:,i]]_-$, which appears as the input of $\text{$k$-th\_largest}(\cdot)$ in (\ref{upper2}). According to the constraint (\ref{robust_1}), we can at most flip $q_{vu}$ bits in $\bm{\hat{X}}_v[u,:]$. Therefore, we pick the top-$q_{vu}$ possible increase to get the upper bound, which accounts for the summation over $k$. Based on the analysis in \cite{robust_1}, the upper bound $\bm{u}_v[i]$ defined in (\ref{upper1}) and (\ref{upper2}) is the tightest under constraints (\ref{robust_1}) and (\ref{robust_2}). Similarly, we can also find the tightest lower bound, $\bm{l}[i]$, for $\bm{\hat{h}_v}[i]$ as
\vspace{-0.5em}
\begin{equation}
\bm{l}_v[i] = (\hat{a}_v\bm{x}_v\bm{\theta})[i]+(\bm{\hat{a}}_v\bm{X}_v\bm{\theta})[i]-\sum_{u\in N(v)}\sum_{k=1}^{q_{vu}} \bm{\hat{a}}_v[u]\bm{\hat{l}}_v[u,i,k], \label{lower1}
\end{equation}
\begin{equation}
\begin{aligned}
\bm{\hat{l}}_v[u,i,k]=\text{$k$-th\_largest}\{(1-\bm{X}_v[u,:])\odot[\bm{\theta}[:,i]]_-\\+\bm{X}_v[u,:]\odot[\bm{\theta}[:,i]]_+\}.\label{lower2}
\end{aligned}
\end{equation}

\begin{figure}
	\vspace{-1em}
	\centering
	\subfigure[ReLU function.]{
		\begin{minipage}[t]{0.45\linewidth}
			\centering
			\includegraphics[width=1\linewidth]{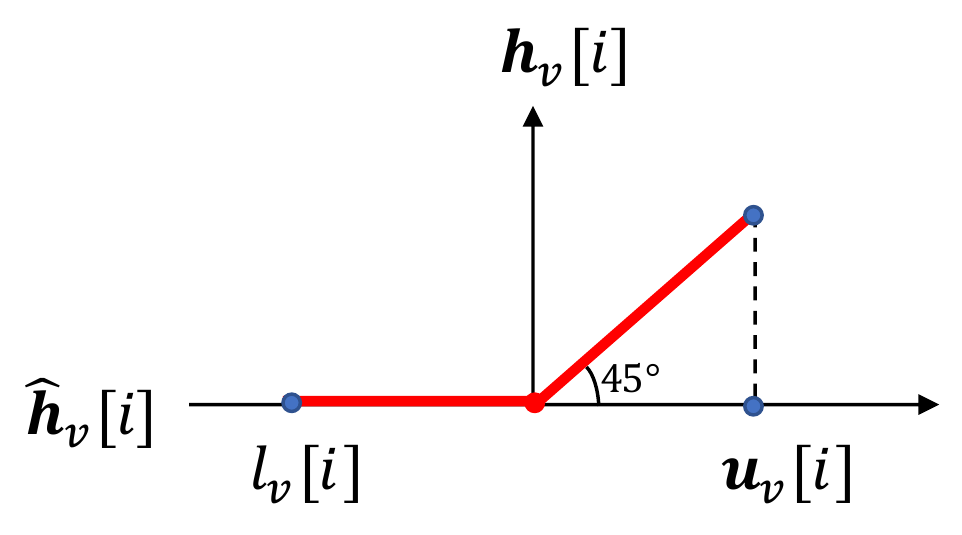}
			\label{fig:relu}
		\end{minipage}
	}
	\subfigure[Convex envelope.]{
		\begin{minipage}[t]{0.45\linewidth}
			\centering
			\includegraphics[width=1\linewidth]{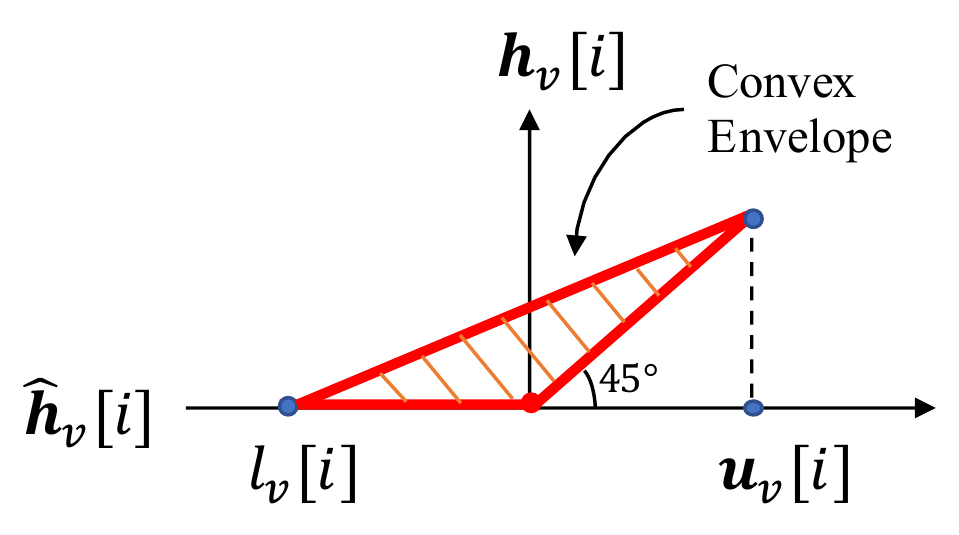}
			\label{fig:envelope}
		\end{minipage}
	}
\vspace{-1em}
	\caption{Illustration of the ReLU function's convex envelope.}
	\label{fig:relu_envelope}
	\vspace{-1.5em}
\end{figure}

After bounding $\bm{\hat{h}}_v[i]$ as $\bm{l}_v[i] \leq \bm{\hat{h}}_v[i] \leq \bm{u}_v[i]$, we can rewrite or approximate the ReLU function in (\ref{relu_h_hat}) based on the signs of $\bm{l}_v[i]$ and $\bm{u}_v[i]$. Specifically, if $\bm{u}_v[i]$ and $\bm{l}_v[i]$ have the same signs, say $0<\bm{l}_v[i]<\bm{u}_v[i]$ or $\bm{l}_v[i]<\bm{u}_v[i]<0$, the ReLU function is reduced into the following linear functions
\begin{equation}
\bm{h}_v[i]=\bm{\hat{h}_v}[i], i \in \mathcal{I}_v^+,
\quad
\bm{h}_v[i]=0, i \in \mathcal{I}_v^-, \label{relu-} 
\end{equation}
where $\mathcal{I}_v^+$ and $\mathcal{I}_v^-$ are the index sets of the entries in $\bm{\hat{h}_v}$ whose upper and lower bounds are both larger or smaller than 0, respectively. On the other hand, if $\bm{l}_v[i]<0<\bm{u}_v[i]$, the ReLU function cannot be directly simplified. To avoid the non-linearity, we can replace it with its convex envelope proposed in \cite{relu_outer} and the operation $\bm{h}_v[i]=\sigma^{(r)}(\bm{\hat{h}_v}[i])$ can be approximated as 
\begin{equation} \label{relu}
\left\{
\begin{array}{lr}
\bm{h}_v[i] \geq  \bm{\hat{h}_v}[i],&  \\
\bm{h}_v[i] \geq 0, & i \in \mathcal{I}_v,\\
\bm{h}_v[i](\bm{u}_v[i]-\bm{l}_v[i]) \leq \bm{u}_v[i](\bm{\hat{h}_v}[i]-\bm{l}_v[i]), &  
\end{array}
\right.
\end{equation}
where $\mathcal{I}_v$ is the index set of the entries in $\bm{\hat{h}_v}$ whose upper and lower bounds differ in signs. A graphical illustration of the ReLU function's convex envelope can be found in Fig. \ref{fig:relu_envelope}.

Based on the linear approximations in (\ref{relu-}) and (\ref{relu}) for the ReLU function, Problem (\ref{Mrobust1}) can be transformed into \vspace{-0.5em}
\begin{equation}\label{Mrobust_relu}
\tilde{z}(\bm{q}_v,\bm{\hat{X}}_v) \triangleq \min_{\{\bm{\tilde{X}}_v,\bm{h}_v,\bm{\hat{h}_v}\}} \hat{c}_v(\bm{h}_v\bm{w}+b),
\end{equation}
subject to (\ref{robust_1}), (\ref{robust_2}),(\ref{relu_h}),  (\ref{relu-}), and (\ref{relu}).

By observation, the feasible region of Problem (\ref{Mrobust1}) is a subset of that of Problem (\ref{Mrobust_relu}), which indicates that the minimum of the objective function in Problem (\ref{Mrobust_relu}) is a lower bound of that in Problem (\ref{Mrobust1}), i.e., $\tilde{z}(\bm{q}_v, \bm{\hat{X}}_v) \leq \hat{z}(\bm{q}_v, \bm{\hat{X}}_v)$.  If $\tilde{z}(\bm{q}_v, \bm{\hat{X}}_v)>0$, then $\hat{z}(\bm{q}_v, \bm{\hat{X}}_v)>0$. Therefore, by solving problem  (\ref{Mrobust_relu}), we can also verify that the predicted label $\hat{c}_v$ is robust if $\tilde{z}(\bm{q}_v,\bm{\hat{X}}_v)>0$.

\subsubsection{Relaxation on Binary Variables}  \label{s5_2_3}
Up to this point, Problem (\ref{Mrobust_relu}) is still intractable because $\bm{\tilde{X}}_v$ is binary (see Eq. (\ref{robust_2})). Therefore, we relax it into the continuous variable between 0 and 1. Then Problem (\ref{Mrobust_relu}) is further relaxed as \vspace{-0.5em}
\begin{equation}\label{Mrobust_bin}
\bar{z}(\bm{q}_v, \bm{\hat{X}}_v) \triangleq  \min_{\{\bm{\tilde{X}}_v,\bm{h}_v,\bm{\hat{h}_v}\}} \hat{c}_v(\bm{h}_v\bm{w}+b),
\end{equation}
subject to (\ref{relu_h}), (\ref{relu-}), (\ref{relu}),
\begin{align}
||\bm{\tilde{X}}_v[u,:]-\bm{\hat{X}}_v[u,:]||_1 \leq q_{vu}, \forall u\in N(v), \label{relax_bin_1}\tag{\theequation a}\\
\bm{\tilde{X}}_v[u,j] \in [0,1], \forall u\in N(v),  j \in \{1, 2, .., p\}, \label{relax_bin_2}\tag{\theequation b}
\end{align}
where $||\cdot||_1$ means the L1 norm of the input vector. It is obvious that Problem (\ref{Mrobust_bin}) is a linear programming problem, and the minimum of whose objective function is a lower bound of that of Problem (\ref{Mrobust_relu}), i.e., $\bar{z}(\bm{q}_v, \bm{\hat{X}}_v) \leq \tilde{z}(\bm{q}_v, \bm{\hat{X}}_v)$. Therefore, the predicted label $\hat{c}_v$ is robust if $\bar{z}(\bm{q}_v,\bm{\hat{X}}_v)>0$. Note that the solution to Problem (\ref{Mrobust_bin}) is not equivalent to the original problem (\ref{Mrobust}). However, we only focus on the sign instead of the value of the objective function. Therefore, the above series of approximation and relaxation meet our needs. By solving Problem (\ref{Mrobust_bin}), we will be able to rule out any non-robust predicted labels and thus guarantee robust predictions of the decentralized GNN binary classifier, which accords with the goal of this paper. In the following, we will solve Problem (\ref{Mrobust_bin}). 

\subsection{Dual Problem and Feasible Solution}  \label{s5_3}
\noindent As mentioned above, Problem (\ref{Mrobust_bin}) is a linear programming problem. Therefore, we can use duality theory to solve it. Specifically, the dual problem of Problem (\ref{Mrobust_bin}) is given by \vspace{-0.5em}
\begin{equation}\label{Mdual}
\begin{aligned}
\max_{\{\bm{\phi}_v,\bm{\beta}_v\}} -\sum_{i=1}^D\bm{\alpha}_v[i](\hat{a}_v\bm{x}_v\bm{\theta})[i]-\text{tr}[\bm{X}_v^T(\bm{\hat{a}}_v^T\bm{\alpha}_v\bm{\theta}^T)]\\-||\bm{\Omega}_v||_1-\sum_{u\in N(v)}q_{vu}\bm{\beta}_v[u]+\text{const}, 
\end{aligned}
\end{equation}
subject to 
\vspace{-1em}
\begin{align}
\bm{\phi}_v \in [0,1]^D, \bm{\beta}_v\in \mathbb{R}^{|N(v)|}, \bm{\beta}_v \succeq 0, \label{dual_1}\tag{\theequation a}
\end{align}
\vspace{-2em}
\begin{align}
\bm{\Omega}_v[u,j]=\max\{\bm{\varepsilon}_v[u,j]-\bm{\beta}_v[u], 0\}, \notag \\ \forall u\in N(v),  j \in \{1, 2, .., p\}, \label{dual_2}\tag{\theequation b}
\end{align}
\vspace{-1.5em}
\begin{align}
&\bm{\varepsilon}_v[u,j]=(1-\bm{\hat{X}}_v[u,j])[(\bm{\hat{a}}_v^T\bm{\alpha}_v\bm{\theta}^T)[u,j]]_++ \notag\\ &\bm{\hat{X}}_v[u,j][(\bm{\hat{a}}_v^T\bm{\alpha}_v\bm{\theta}^T)[u,j]]_-, \forall u\in N(v),  j \in \{1, 2, .., p\}, \label{dual_3}\tag{\theequation c}
\end{align}
\vspace{-1.5em}
\begin{align}
\bm{\alpha}_v[i]=0, \forall i\in \mathcal{I}_v^-,
\quad
\bm{\alpha}_v[i]=(-\hat{c}_v\bm{w}^T)[i], \forall i\in \mathcal{I}_v^+,\label{dual_5}\tag{\theequation d}
\end{align}
\vspace{-1em}
\begin{align}
\bm{\alpha}_v[i]=\frac{\bm{u}_v[i]}{\bm{u}_v[i]-\bm{l}_v[i]}[\hat{c}_v\bm{w}[i]]_--\bm{\phi}_v[i][\hat{c}_v\bm{w}[i]]_+, \forall i \in \mathcal{I}_v, \label{dual_6}\tag{\theequation e}
\end{align}
\vspace{-1em}
\begin{align}
\text{const} = c_vb+ \sum_{i\in \mathcal{I}_v}\frac{\bm{u}_v[i]\bm{l}_v[i]}{\bm{u}_v[i]-\bm{l}_v[i]}[c_v\bm{w}[i]]_-, \label{dual_7}\tag{\theequation f}
\end{align}
where $\text{tr}(\cdot)$ indicates the trace of the input matrix. The proof is available in Appendix A. 

Note that the above dual problem has the similar form to the problems in \cite{robust_1} and  \cite{relu_outer}. Following these two works, we can easily find a good suboptimal solution for Problem (\ref{Mdual}). As mentioned above, since we only focus on whether the objective function in Problem (\ref{Mdual}) is positive, it is not necessary to get its actual value. Therefore, instead of optimizing over all possible values of $\bm{\phi}_v$, we can set $\bm{\phi}_v$ as
\begin{align}
\bm{\phi}_v[i] =\left\{\begin{array}{rcl}
\frac{\bm{u}_v[i]}{\bm{u}_v[i]-\bm{l}_v[i]}, & &\forall i \in \mathcal{I}_v, \\
0,&       &\text{otherwise}.\\
\end{array} \right. \label{feasible_phi}
\end{align}
According to the simulation results in \cite{robust_1}, the choice in (\ref{feasible_phi}) is a good suboptimal solution of the original problem (\ref{Mrobust}) and can efficiently help us verify the robustness of the predicted labels. After the value of $\bm{\phi}_v$ is fixed, $\bm{\alpha}_v$, $\bm{\varepsilon}_v$, and $\text{tr}[\bm{X}_v^T(\bm{\hat{a}}_v^T\bm{\alpha}_v\bm{\theta}^T)]$ all become constant. Then, Problem (\ref{Mdual}) is rewritten as \vspace{-0.5em}
\begin{equation}\label{Mduals}
\max_{\bm{\beta}_v \succeq 0} -||\bm{\Omega}_v||_1-\sum_{u\in N(v)}q_{vu}\bm{\beta}_v[u]+\text{const},
\end{equation}
subject to 
\vspace{-1em}
\begin{align}
\bm{\Omega}_v[u,j]=\max\{\bm{\varepsilon}_v[u,j]-\bm{\beta}_v[u], 0\},\notag\\  \forall u\in N(v),  j \in \{1, 2, .., p\}, \label{duals_2}\tag{\theequation a}
\end{align}
\vspace{-2em}
\begin{align}
\text{const} &= c_vb+ \sum_{i\in \mathcal{I}_v}\frac{\bm{u}_v[i]\bm{l}_v[i]}{\bm{u}_v[i]-\bm{l}_v[i]}[c_v\bm{w}[i]]_-\notag \\&-\sum_{i=1}^D\bm{\alpha}_v[i](\hat{a}_v\bm{x}_v\bm{\theta})[i]-\text{tr}[\bm{X}_v^T(\bm{\hat{a}}_v^T\bm{\alpha}_v\bm{\theta}^T)]. \label{duals_3}\tag{\theequation b}
\end{align}
Again, by using the duality theory, the optimal solution for $\bm{\beta}_v$ is given by 
\begin{equation}
\bm{\beta}_v[u]=\text{$q_{vu}$-th\_largest}(\bm{\varepsilon}_v[u,:]),  \forall u\in N(v),\label{feasible_beta} 
\end{equation}
whose detailed proof can be found in Appendix B. With the closed-form solutions given in (\ref{feasible_phi}) and (\ref{feasible_beta}), the value of the objective function in Problem (\ref{Mdual}) denoted as $\dot{z}(\bm{q}_v, \bm{\hat{X}}_v)$ can be obtained with low complexity. If $\dot{z}(\bm{q}_v, \bm{\hat{X}}_v) >0$, the predicted label $\hat{c}_v$ is robust. 

\section{Robustness Verification in Wireless Communication Systems} \label{s6}
\noindent In the sections above, we have introduced the decentralized GNN binary classifier and formulated a robustness verification problem to verify whether its predicted label is robust. Note that the premise of employing the robustness verification problem is that the local error bound vector, $\bm{q}_v$, is available, which can be estimated from received SINRs and depends on specific communication systems. Therefore, in this section, we study how to obtain $\bm{q}_v$ and justify the robustness of the decentralized GNN binary classifier in uncoded and coded communication systems, respectively.
\vspace{-0.5em}
\subsection{Robustness Verification in Uncoded Communication Systems}\label{s6_1}
\noindent In uncoded communication systems, node $v$ is able to estimate the BER, $\{\epsilon_{vu}, u\in N(v)\}$, of each received signal from its corresponding SINR according to Section \ref{s3_wireless}. Given that the BPSK modulation is adopted, the BER of the received signal $\bm{\hat{x}}_{vu}$ is given by \cite{goldsmith} \vspace{-0.5em}
\begin{equation}
\epsilon_{vu} = Q(\sqrt{2\gamma_{vu}}), \vspace{-0.5em} \label{ber}
\end{equation}
where $Q(\cdot)$ is the Q-function defined as $Q(x)=\int_x^\infty \frac{1}{\sqrt{2\pi}}e^{-x^2/2}dx$\footnote{Our analysis can be generalized to other modulation modes by changing (\ref{ber}) accordingly.}. By i.i.d. assumption for the channel impairment, $q_{vu}$ follows the Binomial distribution, $q_{vu}\sim B(p,\epsilon_{vu})$. On the other hand, the error-tolerance of the decentralized GNN binary classifier is monotonic. Specifically, if it is robust to a received signal with more errors (large $q_{vu}$), it is also robust to a received signal with fewer errors (small $q_{vu}$). Based on the statistical characteristics of $q_{vu}$ and the error-tolerance monotonicity, we propose a new metric, robustness probability, to measure the robustness of the decentralized GNN binary classifier in uncoded communication systems.

As the name implies, the robustness probability of node $v$, denoted as $p_v^{(r)}$, indicates the probability that $\hat{c}_v$ is robust with the received signals $\bm{\hat{X}}_v=\{\bm{\hat{x}}_{vu}, \forall u \in N(v)\}$. To give the formula of $p_v^{(r)}$, we further introduce a new variable $q_v^U$, which is  the solution of the following optimization problem
\vspace{-0.5em}
\begin{equation}\label{Mqu}
q_v^U \triangleq \max_{q} q, 
\end{equation}
\vspace{-1em}
subject to 
\begin{align}
\dot{z}(\bm{q}'_v, \bm{\hat{X}}_v) >0,  \forall q'_{vu}\leq q, u\in N(v), \label{qu_1}\tag{\theequation a}
\end{align}
where $\dot{z}(\cdot, \cdot)$ is defined at the end of Section \ref{s5_3}. Based on the error-tolerance monotonicity, Problem (\ref{Mqu}) can be solved by the bisection algorithm. From the definition of $q_v^U$, if $q_{vu}\leq q_v^U$ holds for all neighbors of node $v$, $\hat{c}_v$ is robust. Therefore, we define $p_v^{(r)}$ as $P(q_{vu}\leq q_v^U, \forall u\in N(v))$. Combined with the fact that  $q_{vu}\sim B(p,\epsilon_{vu})$, the robustness probability of node $v$ is given by \vspace{-0.5em}
\begin{equation}
p_v^{(r)} = \prod_{u\in N(v)}[\sum_{i=0}^{q_v^U}\tbinom{p}{i}(\epsilon_{vu})^i(1-\epsilon_{vu})^{p-i}]. \label{pr} \vspace{-0.5em}
\end{equation}

In practice, a specific robustness requirement is generally imposed on node $v$. In other words, the robustness probability of node $v$ is supposed to be no smaller than a given target robustness probability, $p_v^{(t)}$. If $p_v^{(r)} \geq p_v^{(t)}$, the robustness requirement is satisfied. Otherwise,  retransmission is needed to enhance the prediction robustness, which will be further discussed in Section \ref{s7_1}.

\subsection{Robustness Verification in Coded Communication Systems}\label{s6_2}
\noindent As mentioned in Section \ref{s3_wireless}, in coded communication systems, a transmitted signal can be correctly decoded by the receiver if it is not in outage. Therefore, $\bm{\hat{x}}_{vu}=\bm{x}_u$ and $q_{vu}=0$ if $\log(1+\gamma_{vu})\geq R_u$. Otherwise, the whole transmission packet corresponding to $\bm{x}_u$ is discarded. In this case, we estimate $\bm{\hat{x}}_{vu}$ as a zero vector for the computation of $\hat{c}_v$\footnote{If outages happen, the corresponding signal needs to be estimated, even with random guess, to facilitate the robustness verification process.} and the corresponding maximum number of tolerable errors is $p$, i.e., $q_{vu}=p$. Overall, the value of $q_{vu}$ in coded communication systems is given by 
\begin{align}
q_{vu} =\left\{\begin{array}{rcl}
p, & &\text{if } \log(1+\gamma_{vu})<R_u, \\
0,&       &\text{otherwise}.\\
\end{array} \right. \label{outage_q} \vspace{-1em}
\end{align}
After getting $\bm{q}_v$ based on (\ref{outage_q}), we solve the robustness verification problem (\ref{Mrobust}) and check whether $\dot{z}(\bm{q}_v,\bm{\hat{X}}_v)>0$. If it is, $\hat{c}_v$ is robust and the robustness indicator of node $v$ is set as 1, where the robustness indicator is a binary metric to measure the robustness of the decentralized GNN binary classifier in coded communication systems. Otherwise, the robustness indicator of node $v$ is set as 0 and  retransmission is needed, which will be discussed in Section \ref{s7_2}.

\section{Retransmission Mechanism for Robustness Enhancement} \label{s7}
\noindent When the given robustness requirement is not satisfied, retransmission is needed to enhance the prediction robustness. As mentioned in Section \ref{s3_wireless}, the channel  gain remains static within a symbol block but is i.i.d over different blocks. The channel diversity over different blocks can be exploited by utilizing the maximal-ratio combining (MRC) technique. Specifically, if $\bm{x}_u$ is transmitted for $T$ rounds (one initial transmission and $T-1$ retransmissions) from node $u$ to node $v$, all $T$ received copies of $\bm{x}_u$ can be combined by the MRC technique at node $v$ and a received signal denoted as $\bm{\hat{x}}_{vu}(T)$ is obtained. The effective received SINR of $\bm{\hat{x}}_{vu}(T)$ is \vspace{-0.5em}
\begin{equation}
\gamma_{vu}(T)=\sum_{t=1}^T\gamma_{vu}(t), \label{mrc} \vspace{-0.5em}
\end{equation}
where $\gamma_{vu}(t)$ is the  SINR of the signal received by node $v$ from node $u$ for the $t$-th transmission and is assumed known at node $v$ according to Section \ref{s3_wireless}. Obviously, retransmission can contribute to decreasing the BER and the outage probability for both uncoded and coded communication systems. This further enhances the robustness of the decentralized GNN binary classifier. 

However, traditional retransmission mechanisms aim at reliable transmission and will come to an end only when the target BER is met and no neighbor is in outage for uncoded and coded communication systems, respectively. This will inevitably induce high communication overhead especially for dense networks. To alleviate communication burden of achieving robust predictions, in the following, we adopt the MRC technique to re-design retransmission mechanisms with novel retransmission criteria and stopping rules for the decentralized GNN binary classifier in uncoded and coded communication systems, respectively.

\subsection{Retransmission in Uncoded Communication Systems}\label{s7_1}
\noindent In this part, we develop the retransmission mechanism when the predicted label of node $v$ does not satisfy the robustness requirement, i.e., $p_v^{(r)}<p_v^{(t)}$, in uncoded communication systems.  First, we need to develop a criterion to check whether a neighbor of node $v$ needs to do retransmission. According to (\ref{ber}), node $v$ can estimate the BER $\epsilon_{vu}$ of each received signal based on the received SINR. Generally, when $\epsilon_{vu}$ is  large, the probability of $q_{vu}\leq q_v^U$ will be small, which further leads to a small robustness probability of node $v$.  Obviously, we need to find out those received signals with large BER and ask the corresponding neighbors to do retransmission. To achieve this goal, we first introduce a new variable, $\epsilon_v^U$, which is called BER bound and is the solution to the following  equation \vspace{-1em}
\begin{equation}
\sqrt[|N(v)|]{p_v^{(t)}} = \sum_{i=0}^{q_v^U}\tbinom{p}{i}(\epsilon_v^U)^i(1-\epsilon_v^U)^{p-i}. \label{ber_bound} \vspace{-0.5em}
\end{equation}
Based on the above definition, $\epsilon_v^U$ can be interpreted as the maximum BER of each received signal to guarantee that the robustness probability of $\hat{c}_v$ is no smaller than $p_v^{(t)}$. In this way, we can use $\epsilon_v^U$  as a threshold to check whether retransmission is needed. Specifically, if $\epsilon_{vu}>\epsilon_v^U$, node $u$ needs to retransmit $\bm{x}_u$ to node $v$.

With the above retransmission criterion and the MRC technique, the retransmission mechanism for the decentralized GNN binary classifier in uncoded communication systems is summarized as Algorithm 1 and described as follows. After the initial transmission and prediction, if $p_v^{(r)}<p_v^{(t)}$, node $v$ requests each neighbor $u$ with $\epsilon_{vu}>\epsilon_v^U$ to retransmit their node features. By using the MRC technique, node $v$ coherently combines the received copies for each retransmitted signal, computes their corresponding effective SINR as (\ref{mrc}), and updates the received sliced feature matrix $\hat{\bm{X}}_v$. Then, node $v$ re-predicts its label $\hat{c}_v$ and re-computes the robustness probability $p_v^{(r)}$. The whole transmission process ends when $p_v^{(r)}$ finally becomes no smaller than $p_v^{(t)}$. 
\begin{algorithm}[]
	\small
	\caption{Decentralized GNN Binary Classifier in Uncoded Communication Systems}
	\begin{algorithmic}[1]
		\FOR {node $v$ in $\mathcal{V}$}
		\STATE \textbf{Initial transmission:}
		\STATE Send transmission request to each neighbor in $N(v)$.
		\STATE Receive signals from neighbors: $\{\hat{\bm{x}}_{vu}, u \in N(v)\}$ (or say $\bm{\hat{X}}_v$).
		\STATE Compute BERs for all received signals using (\ref{ber}): $\{\epsilon_{vu}, u \in N(v)\}$. 
		\STATE Use local GNN copy and get the predicted label $\hat{c}_v$.
		\STATE Compute $p_v^{(r)}$  using (\ref{pr}).
		\STATE \textbf{Retransmission:}
		\WHILE {$p_v^{(r)}<p_v^{(t)}$}
		\STATE Get $q_v^U$ by solving Problem (\ref{Mqu}).
		\STATE Get retransmission threshold $\epsilon_v^U$ using (\ref{ber_bound}).
		\STATE Send retransmission request to each neighbor $u$ with $\epsilon_{vu}>\epsilon_v^U$.
		\FOR {neighbor $u$ with $\epsilon_{vu}>\epsilon_v^U$}
		\STATE Retransmit $\bm{x}_u$ to node $v$.
		\ENDFOR
		\STATE Use the MRC technique to combine all received copies for each neighbor with $\epsilon_{vu}>\epsilon_v^U$.
		\STATE Re-compute BERs and update $\bm{\hat{X}}_v$.
		\STATE Re-compute $\hat{c}_v$ and $p_v^{(r)}$.
		\ENDWHILE
		\ENDFOR
	\end{algorithmic}
\end{algorithm}
\subsection{Retransmission in Coded Communication Systems}\label{s7_2}
\noindent In this part, we develop the retransmission mechanism when the predicted label of node $v$ is not robust, i.e., $\dot{z}(\bm{q}_v,\bm{\hat{X}}_v)<0$, in coded communication systems. Obviously, not all received signals but only those in outage need to be retransmitted. In other words, if $\log(1+\gamma_{vu})<R_u$, node $u$  needs to retransmit $\bm{x}_u$ to node $v$. By using the MRC technique, all received copies of $\bm{x}_u$ are coherently combined and the effective SINR is given by (\ref{mrc}), which is used to check whether the effective received signal is in outage. Specifically, after $T$ transmissions, if $\log(1+\gamma_{vu}(T))\geq R_u$, it indicates that $\bm{\hat{x}}_{vu}(T)=\bm{x}_v$ and $q_{vu}(T)=0$. Otherwise, $\bm{\hat{x}}_{vu}(T)=\bm{0}$ and $q_{vu}(T)=p$. In this way, $\bm{\hat{X}}_v$ and $\bm{q}_v$ are updated, which enables node $v$ to re-predict its label and re-solve Problem (\ref{Mrobust}) to check the robustness of the newly-obtained predicted label. If it is robust, the whole transmission process ends. The retransmission mechanism for the decentralized GNN binary classifier in coded communication systems is summarized as Algorithm 2.
\begin{algorithm}[]
	\small
	\caption{Decentralized GNN Binary Classifier in Coded Communication Systems}
	\begin{algorithmic}[1]
		\FOR {node $v$ in $\mathcal{V}$}
		\STATE \textbf{Initial transmission:}
		\STATE Send transmission request to each neighbor in $N(v)$.
		\STATE Receive signals from neighbors: $\{\hat{\bm{x}}_{vu}, u \in N(v)\}$ (or say $\bm{\hat{X}}_v$).
		\STATE Check whether an outage occurs for each received signal and get $\bm{q}_v$ using (\ref{outage_q}).
		\STATE Use local GNN copy and get the predicted label $\hat{c}_v$.
		\STATE Solve Problem (\ref{Mrobust}) and get $\dot{z}(\bm{q}_v,\bm{\hat{X}}_v)$.
		\STATE \textbf{Retransmission:}
		\WHILE {$\dot{z}(\bm{q}_v,\bm{\hat{X}}_v)<0$}
		\STATE Send retransmission request to each neighbor $u$ in outage. 
		\FOR {neighbor $u$ in outage}
		\STATE Retransmit $\bm{x}_u$ to node $v$.
		\ENDFOR
		\STATE Use the MRC technique to combine all received copies for each neighbor in outage.
		\STATE Re-check whether an outage occurs for each newly received signal.
		\STATE Update $\bm{\hat{X}}_v$ and $\bm{q}_v$.
		\STATE Re-compute $\hat{c}_v$ and $\dot{z}(\bm{q}_v,\bm{\hat{X}}_v)$.
		\ENDWHILE
		\ENDFOR
	\end{algorithmic}
\end{algorithm}
\vspace{-1em}

\section{Simulation Results} \label{s8}
\noindent In this section, we use simulation results to show the performance of the decentralized GNN binary classifier in wireless communication systems and validate the effectiveness of the proposed retransmission mechanisms. We also pay attention to the impact of some key parameters about the network design and retransmission mechanisms to shed lights on practical implementation.

\subsection{Simulation Setup}\label{s8_1}
\noindent We conduct the simulation on the synthetic random geometric graph data. To be more specific, we consider a 2,000 m by 2,000 m two-dimensional square area with $N$ nodes, which are uniformly  distributed in the area. Moreover, we set the communication radius as 500 m, which indicates that two nodes are connected if their distance is smaller than 500 m. Furthermore, each graph corresponds to a binary classification problem in a wireless communication system and a specific GNN binary classifier $J(\cdot;\mathcal{W})$ is generated to solve it. Following \cite{simulation}, each entry of the parameters in set $\mathcal{W}$ is independently generated from the Gaussian distribution $\mathcal{N}(0,10^2)$. In a similar way, for all nodes in the graph, each entry of their node features $\{\bm{x}_v\}_{v\in N(v)}$ is independently generated by following the Bernoulli distribution whose successful probability is 0.3. After that, the true labels of all nodes $\{c_v\}_{v\in N(v)}$ are obtained by using the GNN binary classifier in (\ref{C_all}), which suggests that the misclassification rate will be 0 if the wireless transmission is perfect. For simplicity, we assume that all nodes have the same target  robustness probability, and transmit at the same data rate and power, i.e., $p_v^{(t)}=p^{(t)}, R_v=R, P_v=P, \forall v \in \mathcal{V}$. And we do not consider interference among neighbors in the following simulation.  Unless otherwise stated, the values of these three parameters are preset as 80\%, 1 bit/s/Hz, and 0.1 w, respectively. Throughout this section, presented results are the average performance over all nodes in 200 testing graphs. Our simulation parameters are summarized in Table \ref{table:simulation}.
\begin{table}[ht]
	\vspace{-1em}
	\caption{Simulation Parameters}
	\vspace{-1em}
	\scriptsize
	\label{table:simulation}
	\centering
	\begin{tabular}{|c|c|}
		\hline
		Parameter & Value  \\
		\hline
		\hline
		Square Area & 2,000 m $\times$ 2,000 m \\
		\hline
		Communication Radius & 500 m \\
		\hline
		Bandwidth & 10 MHz \\
		\hline
		Noise Spectral Density & -174 dBm/Hz\\
		\hline
		Path Loss Model& 128.1+37.6log(d[km])\\
		\hline
		Shadowing Standard Deviation, $std$ & 8 dB \\
		\hline
		\tabincell{c}{Target Robustness Probability, $p^{(t)}$} &  80\%\\
		\hline
		\tabincell{c}{Transmit Data Rate, $R$} & 1 bit/s/Hz\\
		\hline
		\tabincell{c}{Transmit Power, $P$} & 0.1 w\\
		\hline
		\tabincell{c}{Node Feature Dimension, $p$} & 32\\
		\hline
		\tabincell{c}{Hidden State Dimension, $D$} & 32\\
		\hline
		Graph Filter, $\bm{\hat{A}}$ & 	\tabincell{c}{Unnormalized graph filter,\\ $\bm{\hat{A}}=\bm{A}+\bm{I}_N$\cite{unnormal}}\\
		\hline
	\end{tabular}
	\vspace{-1.5em}
\end{table}

\subsection{Simulation Results in Uncoded Communication Systems}\label{s8_2}
\subsubsection{The Performance without Retransmission}\label{s8_2_1}
\begin{figure}
	\vspace{-2em}
	\centering
	\subfigure[\scriptsize{Misclassification rate w/o and w/ retransmission.}]{
		\begin{minipage}[t]{0.9\linewidth}
			\centering
			\includegraphics[width=0.8\linewidth]{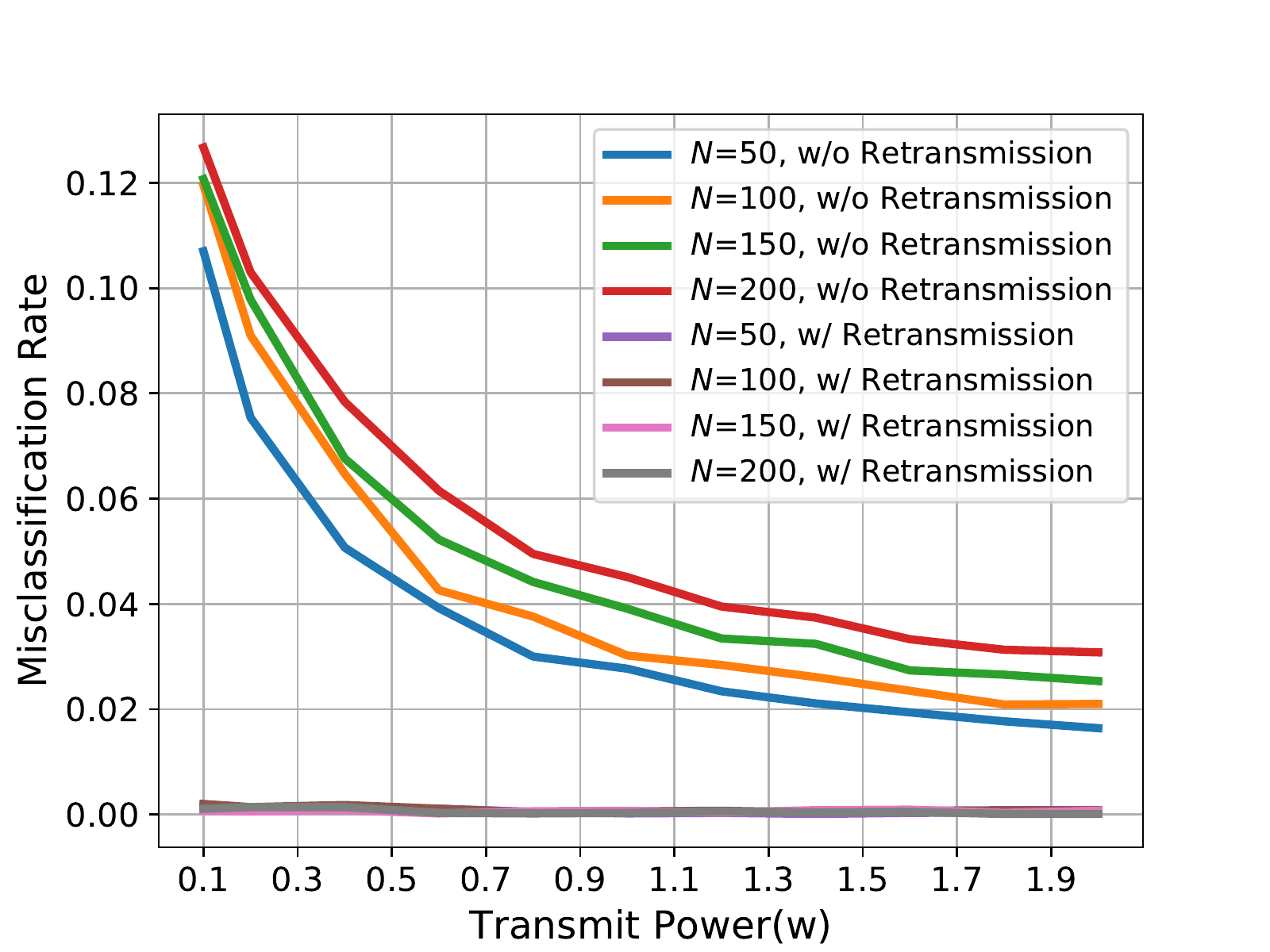}
			\vspace{-1em}
			\label{fig:bit_node_miss}
		\end{minipage}
	}
	\subfigure[\scriptsize{Percentage of robust nodes w/o retransmission.}]{
		\begin{minipage}[t]{0.9\linewidth}
			\centering
			\includegraphics[width=0.8\linewidth]{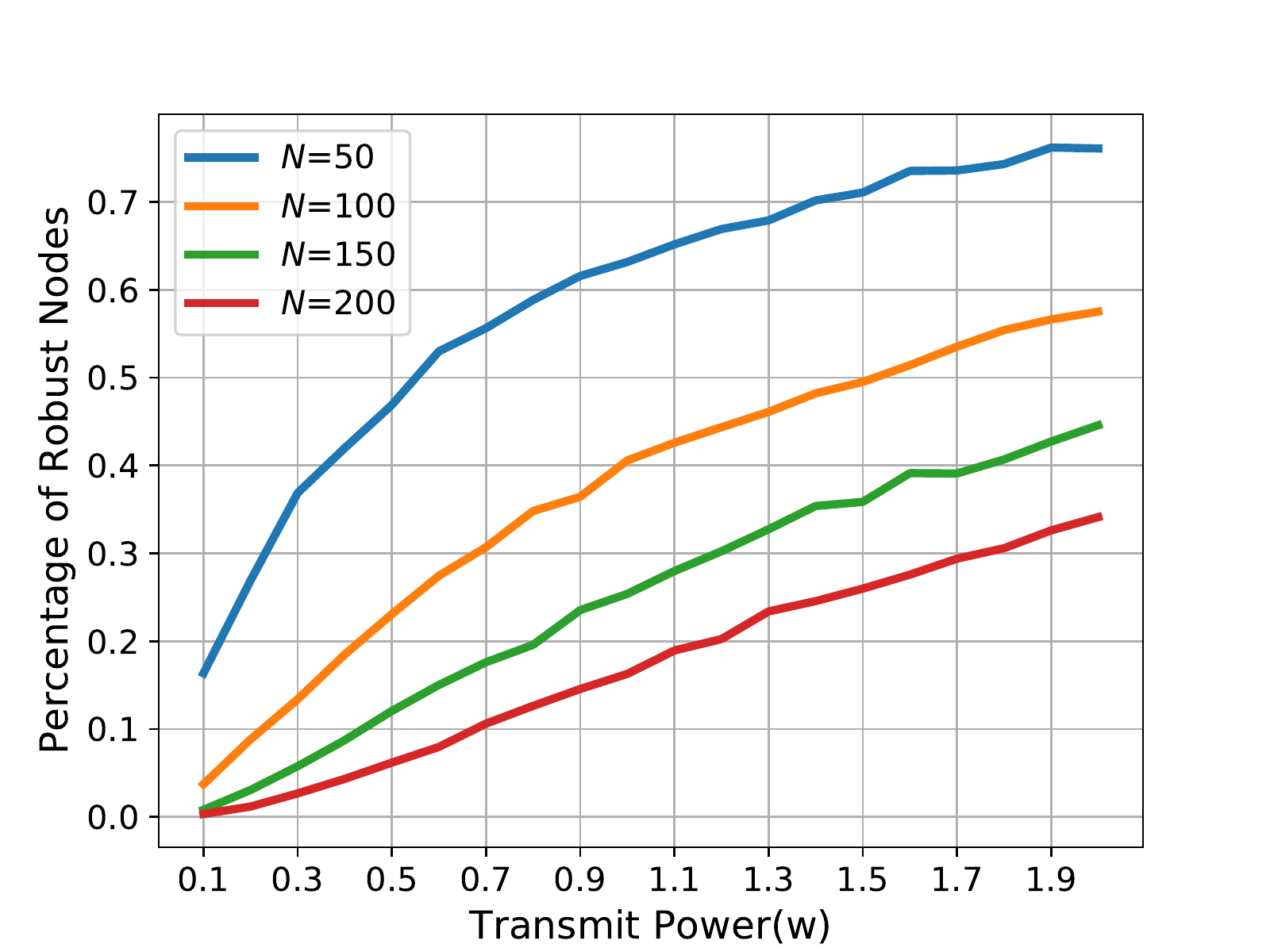}
			\vspace{-1em}
			\label{fig:bit_node_robust}
		\end{minipage}
	}
	\subfigure[\scriptsize{Average needed transmission rounds.}]{
		\begin{minipage}[t]{0.9\linewidth}
			\centering
			\includegraphics[width=0.8\linewidth]{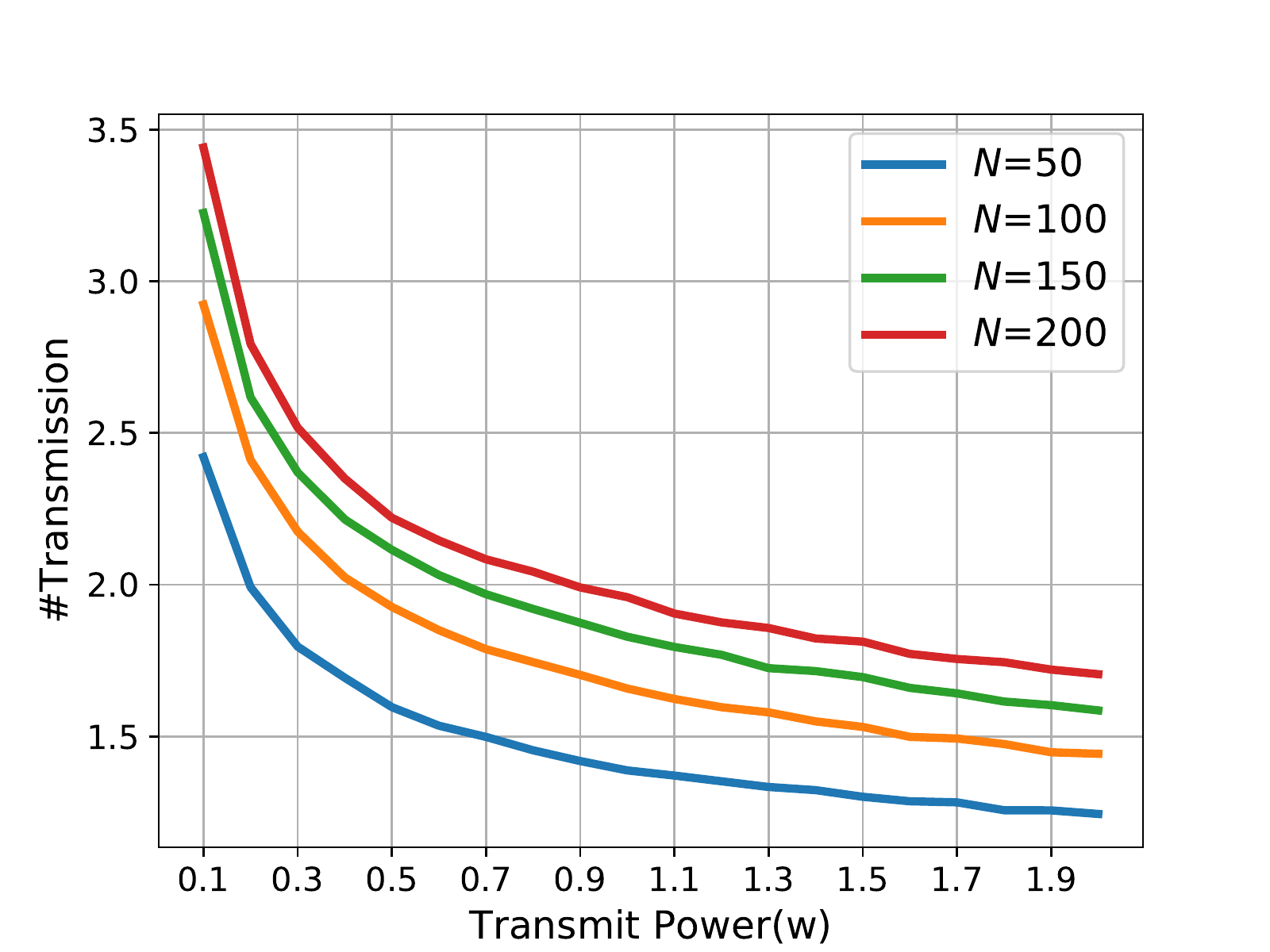}
			\vspace{-1em}
			\label{fig:bit_node_trans}
		\end{minipage}
	}
	\vspace{-0.5em}
	\caption{Simulation results for  the decentralized GNN binary classifier in uncoded communication systems.}
	\label{fig:bit_node}
	\vspace{-2em}
\end{figure}

First, we test the performance of the decentralized GNN binary classifier in uncoded communication systems without retransmission. Specifically, we pay attention to the misclassification rate and the percentage of robust nodes without retransmission. The results are summarized in Figs. \ref{fig:bit_node_miss} and \ref{fig:bit_node_robust}, where ``w/o" is the abbreviation for ``without".  From the figures, it can be seen that when the transmit power is small, the imperfect transmission would lead to about 11\%-13\% incorrect predictions. With the increase of the transmit power, the received SINR increases, which further decreases the BER of the received signal. Therefore, the misclassification rate decreases and the percentage of robust nodes increases. Moreover, both metrics change at a decreasing speed when the transmit power increases, indicating that increasing power only leads to a limited performance enhancement for the decentralized GNN binary classifier. Without other mechanisms, it takes extremely high power consumption to achieve the robustness requirement, which suggests the necessity of the retransmission mechanism especially for power-limited systems.

\begin{table}
	\vspace{-1em}
	\scriptsize
	\caption{\#Transmission Ratio between the Traditional and the Proposed Retransmission Mechanism for Scenarios with $N=200$}
	\vspace{-1em}
	\label{trans_gain}
	\centering
	\begin{tabular}{|c|c|c|c|c|c|}
		\hline
		\diagbox{\tabincell{c}{Communication \\Mode}}{$Ratio_T$}{\tabincell{c}{Transmit Power\\(w)}}& 0.1 & 0.5 & 1.0 & 1.5 &2.0\\
		\hline
		Uncoded & 1.52  &1.47&  1.42&  1.42&  1.42 \\
		\hline
		Coded& 2.02 &2.00& 1.94 & 1.99& 1.59  \\
		\hline
	\end{tabular}
\end{table}

\subsubsection{The Performance with Retransmission}\label{s8_2_2}
To show the performance of the decentralized GNN in uncoded communication systems with retransmission, we focus on the misclassification rate with retransmission and the number of average needed transmission rounds to achieve the robustness requirement. The results are summarized in Figs. \ref{fig:bit_node_miss} and \ref{fig:bit_node_trans},  where ``w/" is the abbreviation for ``with". From Fig. \ref{fig:bit_node_miss}, after each node achieving the preset robustness requirement $p^{(t)}=80\%$, the misclassification rate is around 0, which suggests that the retransmission mechanism proposed in Section \ref{s7_1} can efficiently decrease incorrect predictions and enhance the prediction robustness. Furthermore, Fig. \ref{fig:bit_node_trans} suggests that the number of average needed transmission rounds to achieve the robustness requirement decreases with the transmit power. The reason has been discussed in Section \ref{s8_2_1}. Larger transmit powers bring about fewer errors, and therefore fewer transmission rounds are needed to achieve the preset $p^{(t)}$. Undoubtedly, both metrics are closely related to the value of $p^{(t)}$, which will be further discussed later.

To further validate the advantage of the proposed retransmission mechanism, we compare it with the traditional retransmission mechanism where the retransmission will stop until the BER of the received signals from all neighbors are smaller than a given threshold, which is manually set in advance. To achieve a fair comparison, the BER threshold of the traditional retransmission mechanism should be set to be comparable with the proposed retransmission mechanism, i.e., $\epsilon_v^U$. However the value of $\epsilon_v^U$ changes for each transmission. We notice that  its value varies from $3\times10^{-4}$ to $9\times10^{-5}$  during the above simulation. Therefore, to be conservative about our performance gain, we set the BER threshold of the traditional retransmission mechanism as $3\times10^{-4}$ in the following simulation. Specifically, we pay attention to the gain in terms of the number of average needed transmission rounds, \#Transmission, and the results are summarized in Table \ref{trans_gain}, where $Ratio_T$ refers to the ratio between the \#Transmission of the traditional and the proposed retransmission mechanisms.  The results suggest that our proposed retransmission mechanism is superior  to the traditional retransmission mechanism especially when the transmit power is small. 

\subsubsection{The Impact of Target Robustness Probability} \label{s8_2_3}
\begin{figure}
	\vspace{-2.5em}
	\centering
	\subfigure[\scriptsize{Misclassification rate w/ retransmission.}]{
		\begin{minipage}[t]{0.9\linewidth}
			\centering
			\includegraphics[width=0.8\linewidth]{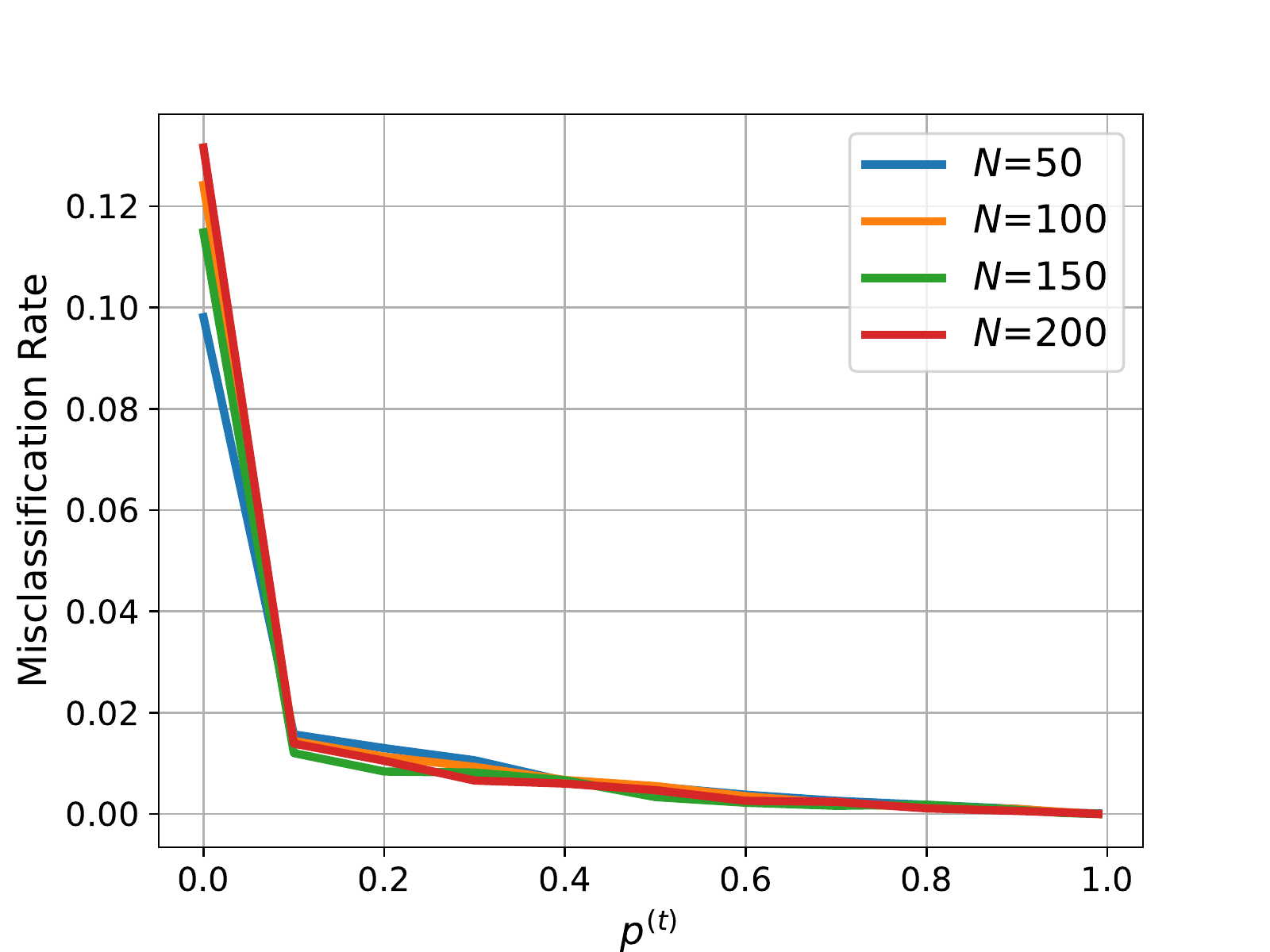}
			\vspace{-1em}
			\label{fig:bit_pt_miss}
		\end{minipage}
	}
	\subfigure[\scriptsize{Percentage of robust nodes w/o retransmission.}]{
		\begin{minipage}[t]{0.9\linewidth}
			\centering
			\includegraphics[width=0.8\linewidth]{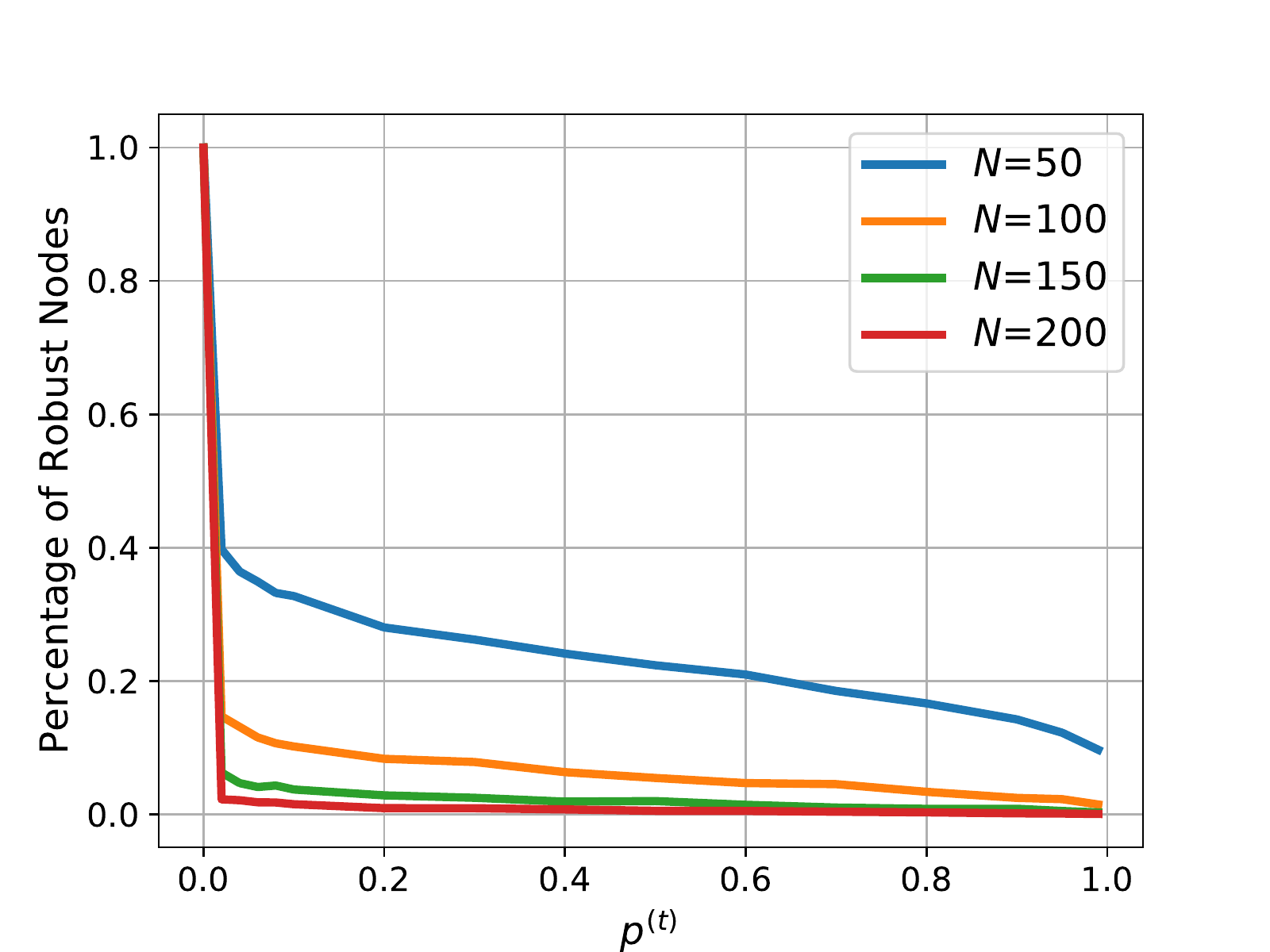}
			\vspace{-1em}
			\label{fig:bit_pt_robust}
		\end{minipage}
	}
	\subfigure[\scriptsize{Average needed transmission rounds.}]{
		\begin{minipage}[t]{0.9\linewidth}
			\centering
			\includegraphics[width=0.8\linewidth]{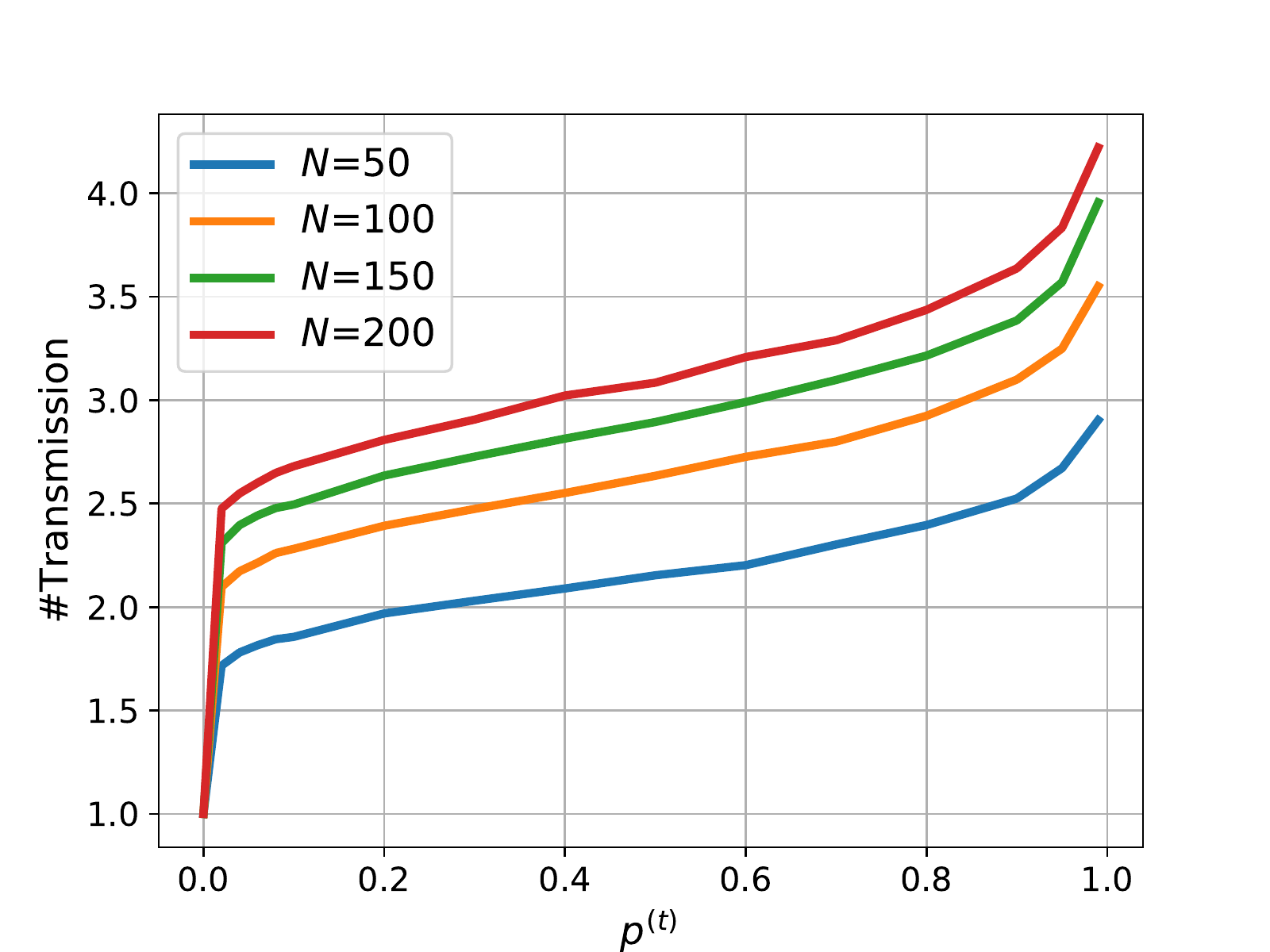}
			\vspace{-1em}
			\label{fig:bit_pt_trans}
		\end{minipage}
	}
	\vspace{-0.5em}
	\caption{Simulation results in uncoded communication systems with different target robustness probability.}
	\label{fig:bit_pt}
	\vspace{-1em}
\end{figure}

In uncoded communication systems, the performance highly depends on the value of the target robustness probability as mentioned above. Therefore, in this part, we study the impact of $p^{(t)}$ with fixed transmit power $P=0.1$ w and the simulation results are summarized in Fig. \ref{fig:bit_pt}. Obviously, with the increase of $p^{(t)}$, the misclassification rate with retransmission decreases and finally converges to 0 (for $p^{(t)}\geq 80\%$) as suggested in Fig. \ref{fig:bit_pt_miss}. Also, fewer nodes can achieve the robustness requirement without retransmission  as indicated by Fig. \ref{fig:bit_pt_robust} and eventually more transmission rounds are needed as shown in Fig. \ref{fig:bit_pt_trans}. Furthermore, it is observed that all  three metrics change very sharply when increasing $p^{(t)}$ from 0 to $10\%$, which indicates that a low robustness requirement already helps a lot to enhance the prediction accuracy. Another important observation is that the number of average needed transmission rounds increases very quickly when $p^{(t)} \geq 90\%$. Given that the misclassification rate converges to 0 when $p^{(t)}$ reaches 80\%,  there is no need to set $p^{(t)}$ larger than 80\% for robust predictions, which does not enhance the prediction accuracy but leads to extra communication overhead. Obviously, the threshold value (80\% in this test setting) varies with different applications, but the key insight concerning the trade-off between prediction accuracy and communication overhead when selecting the value of $p^{(t)}$  always holds. Indeed, it should be reasonably chosen according to the specific accuracy requirement in practice.

\subsection{Simulation Results in Coded Communication Systems}\label{s8_3}
\subsubsection{The Performance without Retransmission}\label{s8_3_1}
\begin{figure}
	\vspace{-2.5em}
	\centering
	\subfigure[\scriptsize{Misclassification rate w/o retransmission.}]{
		\begin{minipage}[t]{0.9\linewidth}
			\centering
			\includegraphics[width=0.8\linewidth]{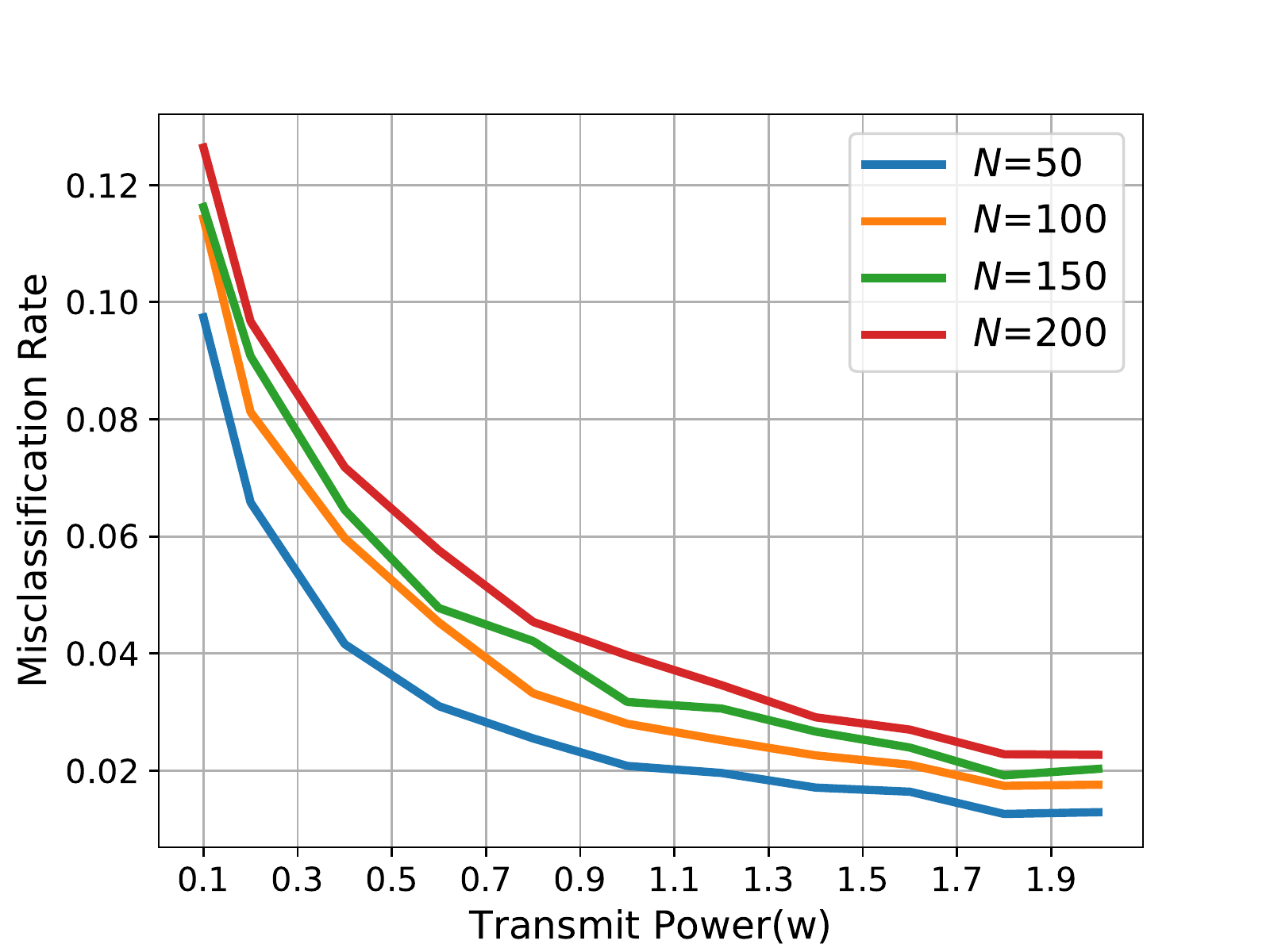}
			\vspace{-1em}
			\label{fig:fading_node_miss}
		\end{minipage}
	}
	\subfigure[\scriptsize{Percentage of robust nodes w/o retransmission.}]{
		\begin{minipage}[t]{0.9\linewidth}
			\centering
			\includegraphics[width=0.8\linewidth]{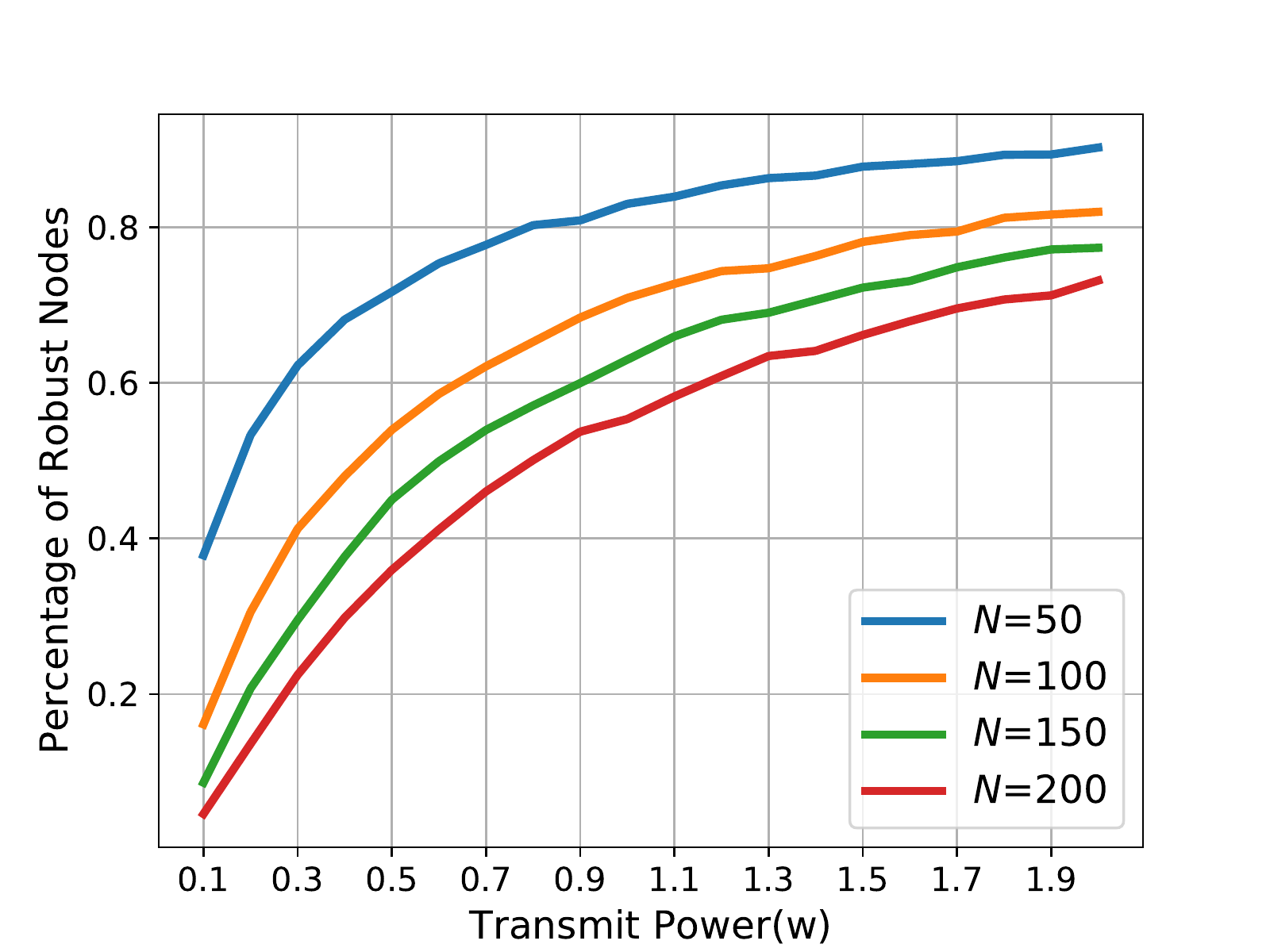}
			\vspace{-1em}
			\label{fig:fading_node_robust}
		\end{minipage}
	}
	\subfigure[\scriptsize{Average needed transmission rounds.}]{
		\begin{minipage}[t]{0.9\linewidth}
			\centering
			\includegraphics[width=0.8\linewidth]{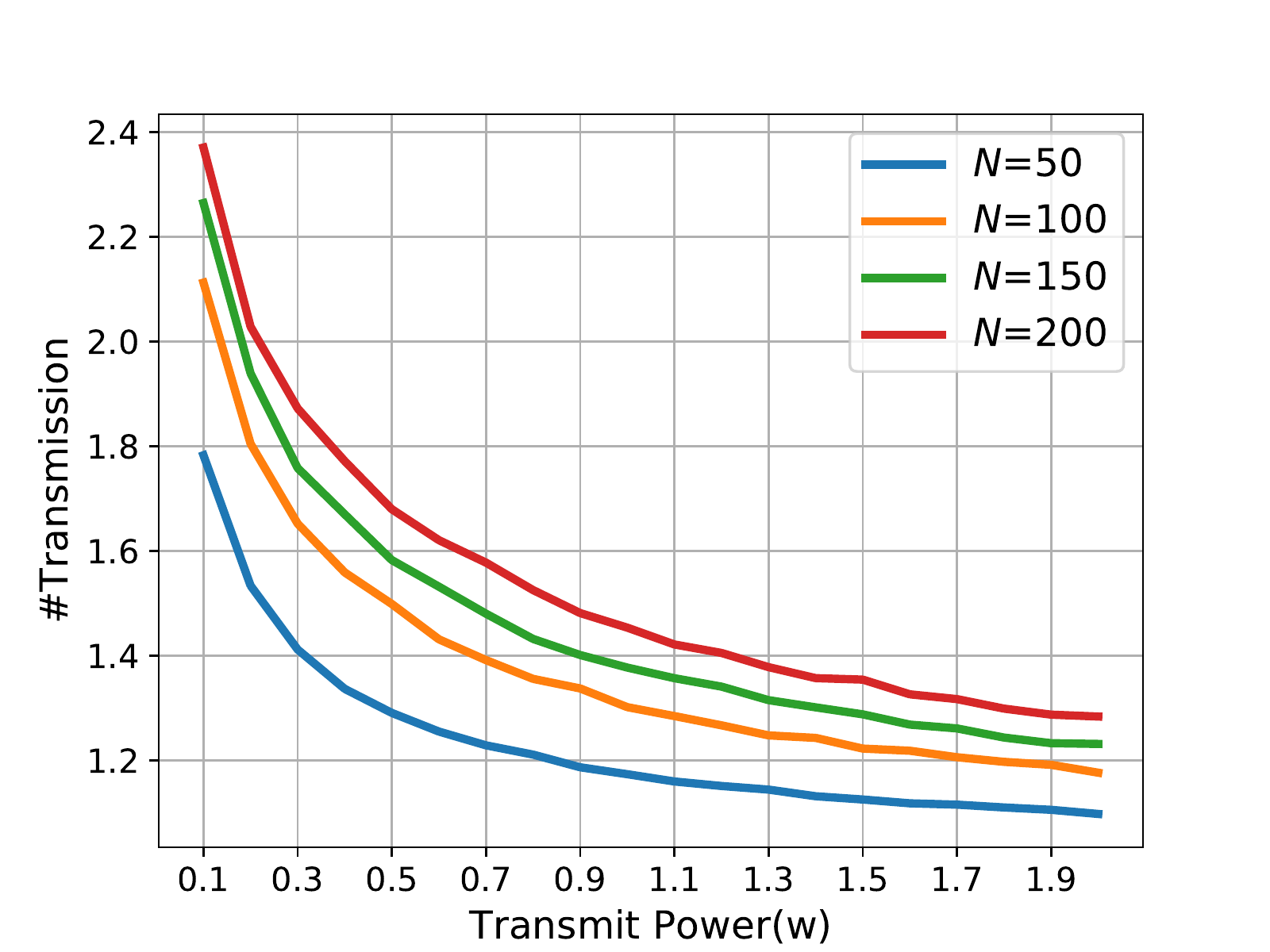}
			\vspace{-1em}
			\label{fig:fading_node_trans}
		\end{minipage}
	}
	\vspace{-0.5em}
	\caption{Simulation results for decentralized GNN binary classifier in coded communication systems.}
	\label{fig:fading_node}
	\vspace{-2em}
\end{figure}

Similar to Section \ref{s8_2_1}, we use the misclassification rate and the percentage of robust nodes without retransmission as two metrics to show the performance of the decentralized GNN binary classifier in coded communication systems  without retransmission. The results are summarized in Figs. \ref{fig:fading_node_miss} and \ref{fig:fading_node_robust}. From the figures, it can be seen that the impact of the transmit power is similar to that in uncoded  systems. However, by comparing the simulation results in Figs. \ref{fig:bit_node} and \ref{fig:fading_node}, with the same transmit power, the decentralized GNN in coded communication systems has a smaller misclassification rate and a larger percentage of robust nodes, which indicates the benefits of coding before transmission.

\subsubsection{The Performance with Retransmission}\label{s8_3_2}
In this part, we study the performance of the decentralized GNN binary classifier in coded systems with retransmission.  Different from uncoded communication systems where retransmission can only guarantee that the predicted label is robust with a probability $p^{(t)}$, the predicted label in coded communication systems with retransmission is definitely robust, i.e., equals the true label. This indicates that the misclassification rate with retransmission is always 0 in coded communication systems. Therefore, we only use the number of average needed transmission rounds to achieve robustness as a metric in this part and the results are summarized in Fig. \ref{fig:fading_node_trans}. Meanwhile, the comparison with the traditional retransmission mechanism where the retransmission will stop until no neighbor is in outage is presented in Table \ref{trans_gain}.

Similar to the analysis in Section \ref{s8_2_2}, the number of average needed transmission rounds decreases with the transmit power and is only 50\% of that with the traditional retransmission mechanism. Moreover, with the same transmit power, coded communication systems need fewer transmission rounds to achieve robustness requirements than  uncoded communication systems. This again validates the superiority of coded systems in terms of prediction robustness.

\subsubsection{The Impact of Transmit Data Rate} \label{s8_3_3}
\begin{figure}
	\vspace{-3em}
	\centering
	\subfigure[\scriptsize{Misclassification rate w/o retransmission.}]{
		\begin{minipage}[t]{0.9\linewidth}
			\centering
			\includegraphics[width=0.8\linewidth]{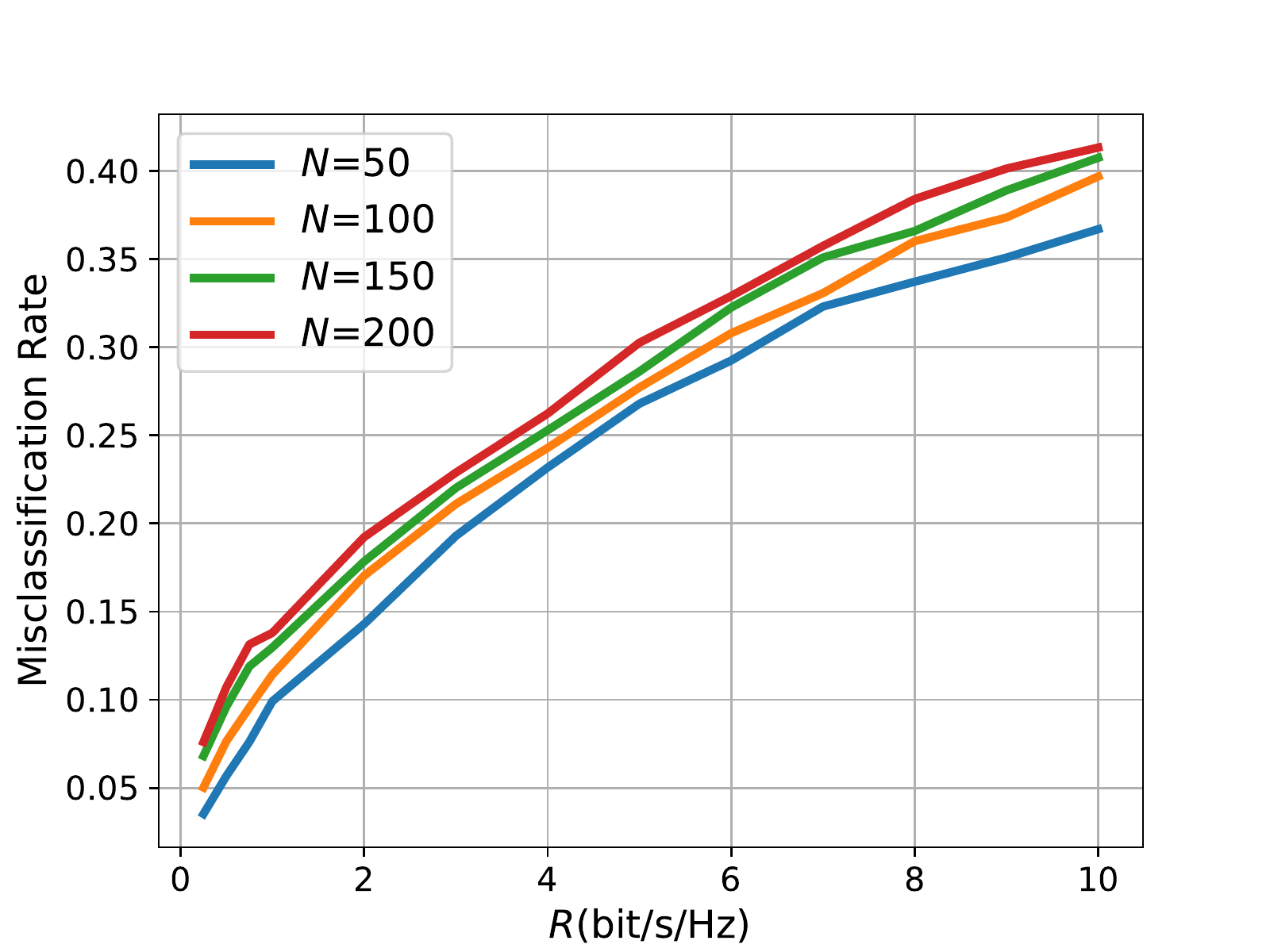}
			\vspace{-4em}
			\label{fig:dr_node_miss}
		\end{minipage}
	}
	\subfigure[\scriptsize{Percentage of robust nodes w/o retransmission.}]{
		\begin{minipage}[t]{0.9\linewidth}
			\centering
			\includegraphics[width=0.8\linewidth]{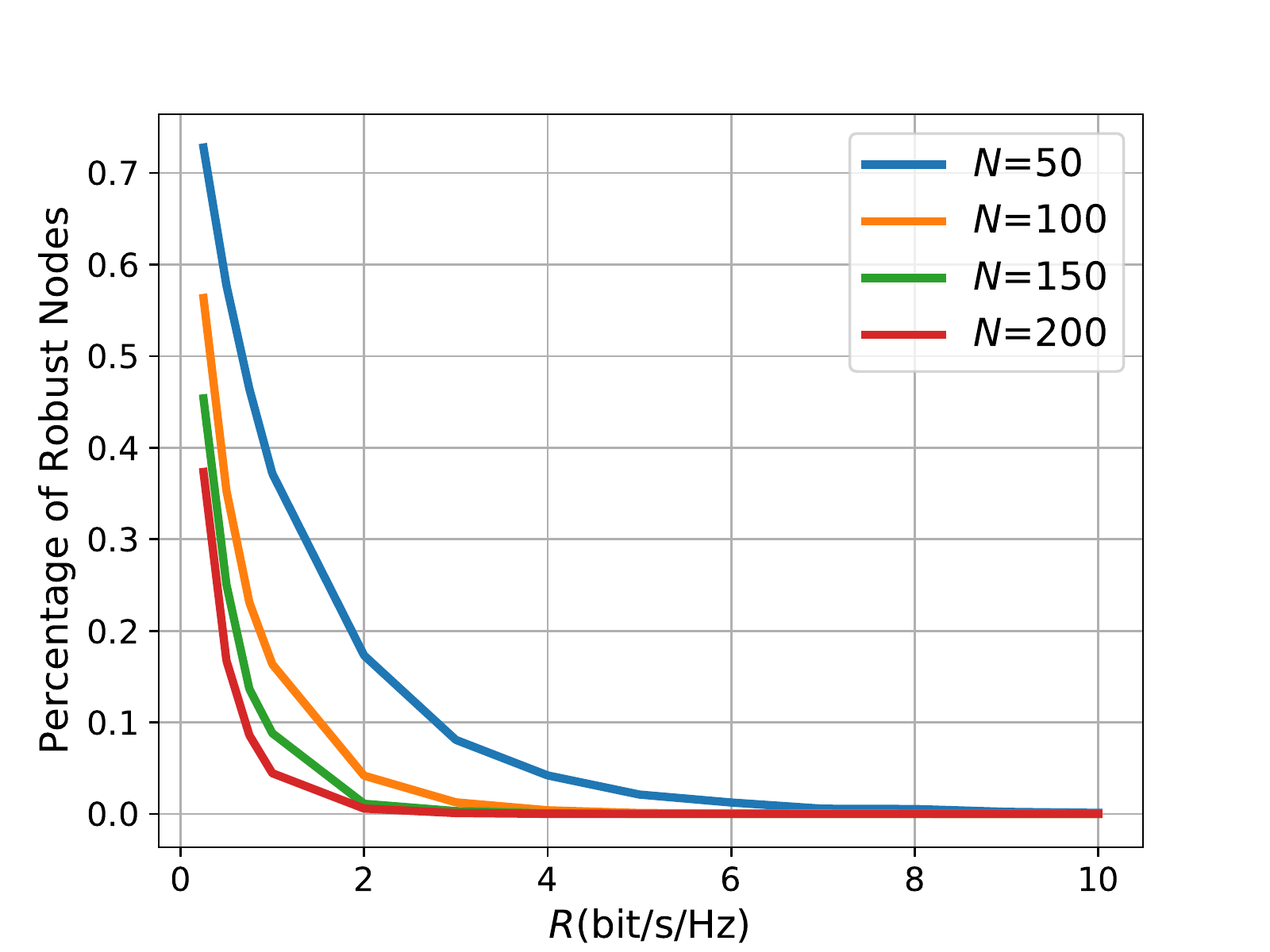}
			\vspace{-4em}
			\label{fig:dr_node_robust}
		\end{minipage}
	}
	\subfigure[\scriptsize{Average needed transmission rounds.}]{
		\begin{minipage}[t]{0.9\linewidth}
			\centering
			\includegraphics[width=0.8\linewidth]{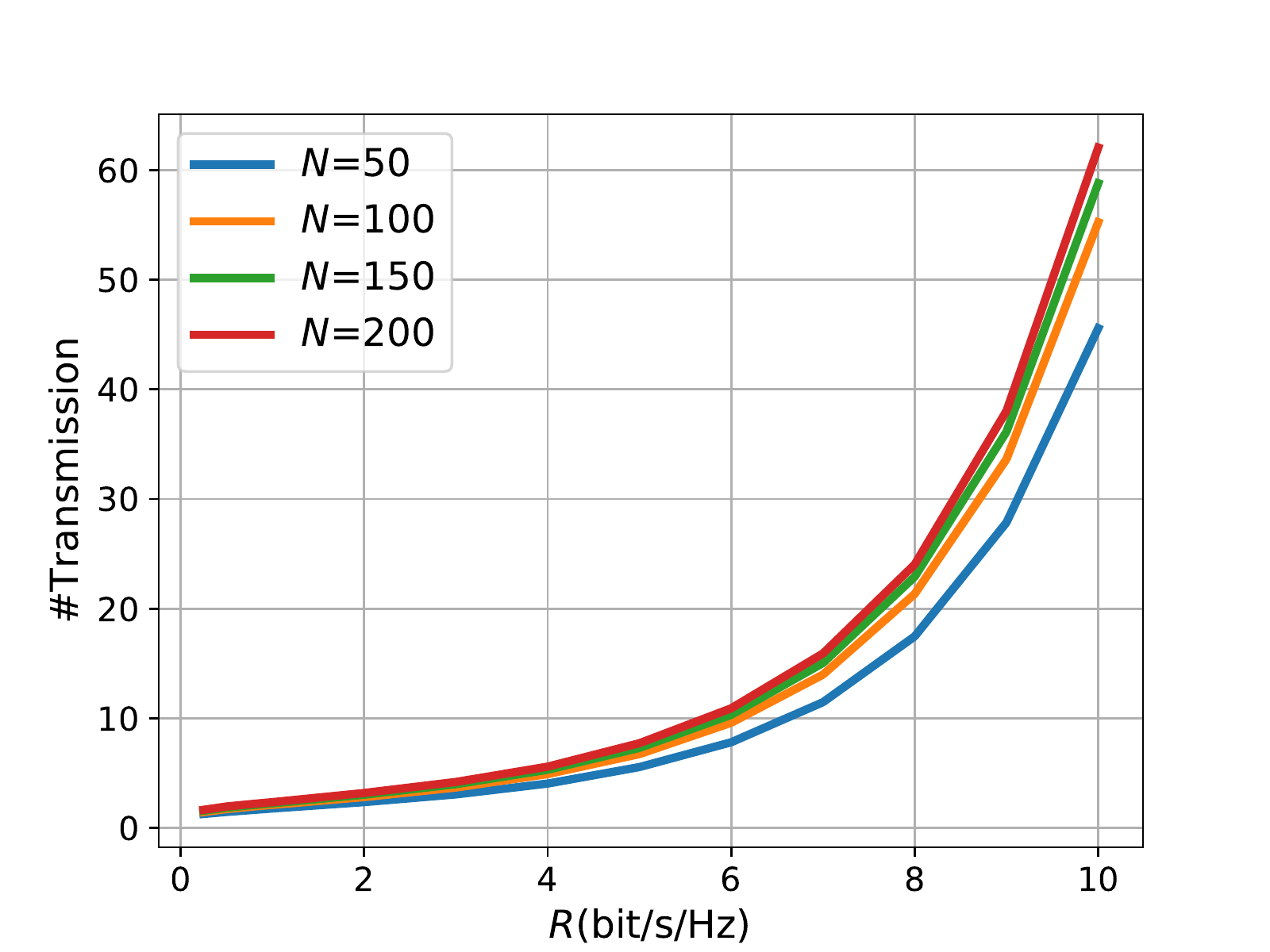}
			\vspace{-4em}
			\label{fig:dr_node_trans}
		\end{minipage}
	}
	\subfigure[\scriptsize{Effective data rate per transmission round.}]{
		\begin{minipage}[t]{0.9\linewidth}
			\centering
			\includegraphics[width=0.8\linewidth]{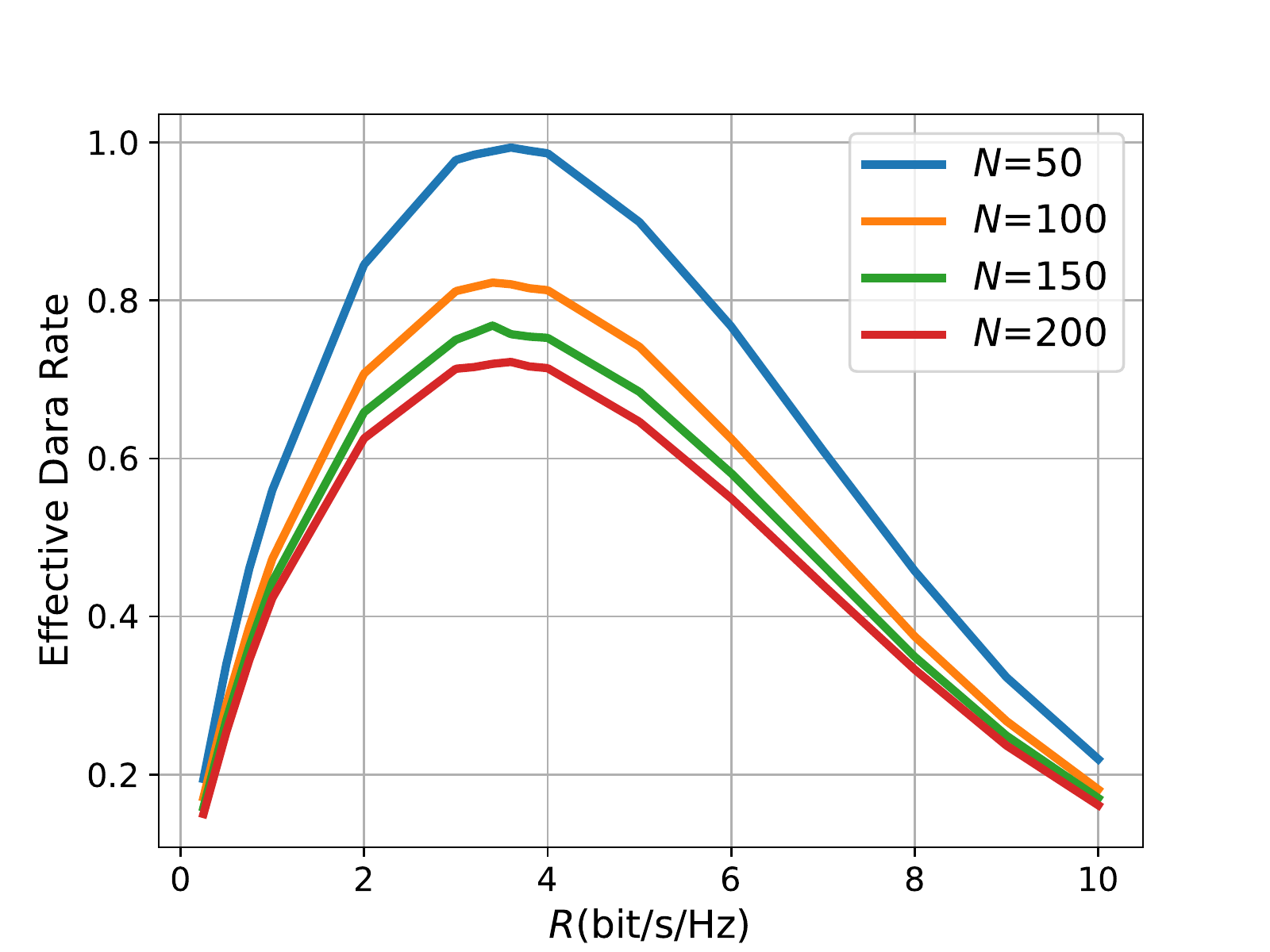}
			\vspace{-4em}
			\label{fig:dr_node_delay}
		\end{minipage}
	}
	\vspace{-1em}
	\caption{Simulation results in coded communication systems with different transmit data rates.}
	\label{fig:dr_node}
	\vspace{-2em}
\end{figure}

Transmit data rate is an important parameter in coded communication systems, which directly influences the outage probability and the  time consumption. Therefore, we study the impact of transmit data rate on the robustness of coded systems. The results are summarized in Fig. \ref{fig:dr_node}, where the effective data rate per transmission round in Fig. \ref{fig:dr_node_delay} is defined as the ratio between the transmit data rate and the number of needed transmission rounds to achieve robustness. Obviously, it is negatively related to the time consumption of the decentralized GNN binary classifier to achieve robustness (Algorithm 2). 

From Figs. \ref{fig:dr_node_miss} and \ref{fig:dr_node_robust}, larger transmit data rate would deteriorate the prediction robustness (larger misclassification rates and smaller percentages of robust nodes without retransmission). With the same preset transmit power (0.1 w), larger transmit data rate would lead to higher outage probability of received signals. Accordingly, more transmission rounds are needed to achieve robustness with larger transmit data rate as shown in Fig. \ref{fig:dr_node_trans}. Furthermore, the effective transmit data rate per transmission round first increases and then decreases with the transmit data rate as can be seen in Fig. \ref{fig:dr_node_delay}. From this figure, a suitable value of $R$ (that maximizes the effective data rate) should be 3.5 bit/s/Hz for this test setting, which is variant in different applications. The results indicate that the selection of the transmit data rate also involves various trade-offs (similar to the conclusion in Section \ref{s8_2_3} about $p^{(t)}$), which should be reasonably chosen for practical implementation.

\subsection{The Impact of Network Parameters} \label{s8_4}
\noindent To get more insights about the wireless network design and practical implementation, we study the impact of two important network parameters, i.e., the node density and the node feature dimension, in this section.
\subsubsection{The Impact of Node Density} \label{s8_4_1}
First, we study the impact of the node density on the decentralized GNN binary classifier in wireless communication systems. To this end, we conduct simulation on graphs with different numbers of nodes located in a fixed area size (2,000m $\times$ 2,000m). The results are summarized in Figs. \ref{fig:bit_node}, \ref{fig:bit_pt}, \ref{fig:fading_node}, and \ref{fig:dr_node}. From these figures, increasing the node density would deteriorate the performance of the decentralized GNN binary classifier. Specifically, in both uncoded and coded communication systems, larger node densities lead to larger misclassification rates and smaller percentages of robust nodes without retransmission.  When retransmission mechanism is used, the predicted results must satisfy the target robustness probability. Therefore, the misclassification rate with retransmission are always around 0 regardless of the node density and the corresponding curves are close to each other. However the corresponding communication overhead is different. Specifically, more transmission rounds are needed for networks with larger node densities to achieve a given robustness requirement. This can be explained as follows. The node density directly influences the number of neighbors of each node in the graph, i.e., node degree $|N(v)|$. As shown in Table \ref{degree}, larger node densities correspond to larger node degrees. For uncoded communication systems, larger node degrees lead to smaller robustness probability for given SINR or BER of the received signal based on (\ref{pr}). On the other hand, for coded communication systems, larger node degrees may increase the number of neighbors in outage, which deteriorates the prediction robustness. Overall, networks with small node densities are  preferred in terms of prediction robustness. Nowadays, however, ultra-dense networks (UDNs) are prevalent, where neighborhood sampling technique can be utilized to enhance the prediction robustness of the decentralized GNN.
\begin{table}
	\vspace{-1em}
	\scriptsize
	\caption{Average Node Degree for Graphs with Different Node Densities}
	\vspace{-1em}
	\label{degree}
	\centering
	\begin{tabular}{|c|c|c|c|c|}
		\hline
		Number of Nodes, $N$& 50 & 100 & 150 & 200\\
		\hline
		Average Node Degree & 7.61 &15.56 &23.33 &31.15\\
		\hline
	\end{tabular}
\end{table}

\subsubsection{The Impact of Node Feature Dimension} \label{s8_4_2}
\begin{figure}
	\vspace{-1em}
	\centering
	\subfigure[\scriptsize{Misclassification rate w/o and w/ retransmission.}]{
		\begin{minipage}[t]{0.9\linewidth}
			\centering
			\includegraphics[width=0.8\linewidth]{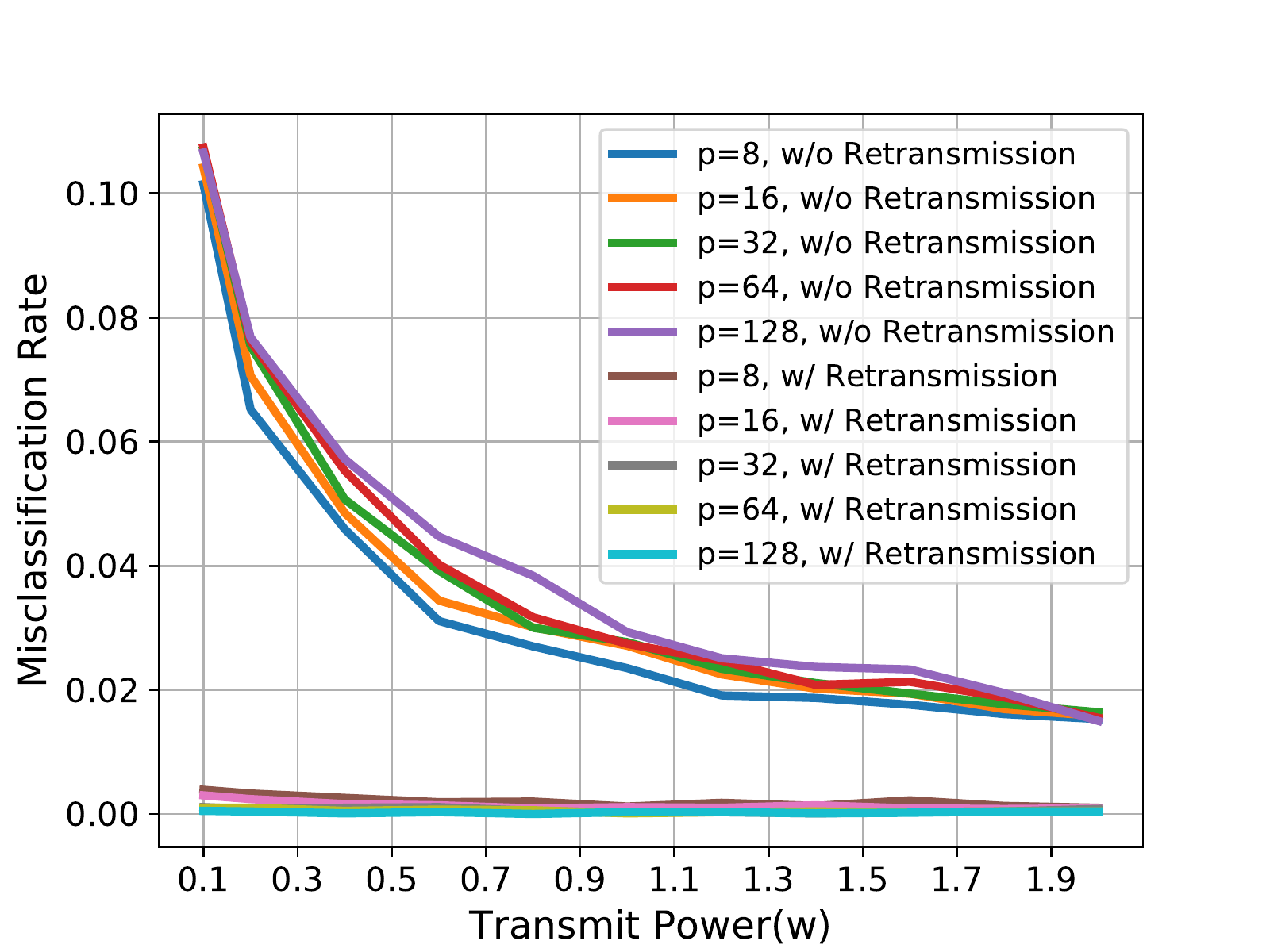}
			\vspace{-1em}
			\label{fig:bit_p_miss}
		\end{minipage}
	}
	\subfigure[\scriptsize{Percentage of robust nodes w/o retransmission.}]{
		\begin{minipage}[t]{0.9\linewidth}
			\centering
			\includegraphics[width=0.8\linewidth]{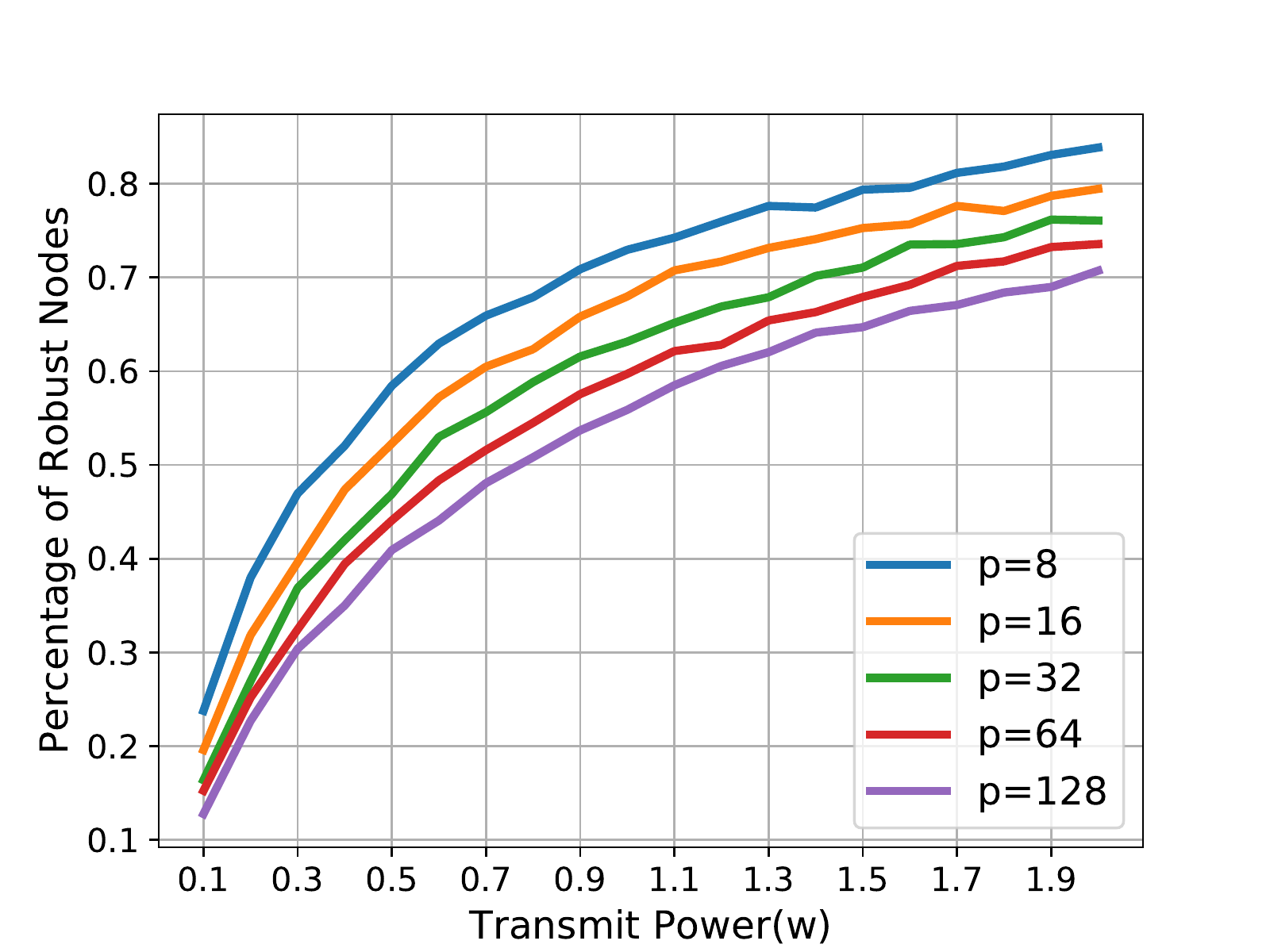}
			\vspace{-1em}
			\label{fig:bit_p_robust}
		\end{minipage}
	}
	\subfigure[\scriptsize{Average needed transmission rounds.}]{
		\begin{minipage}[t]{0.9\linewidth}
			\centering
			\includegraphics[width=0.8\linewidth]{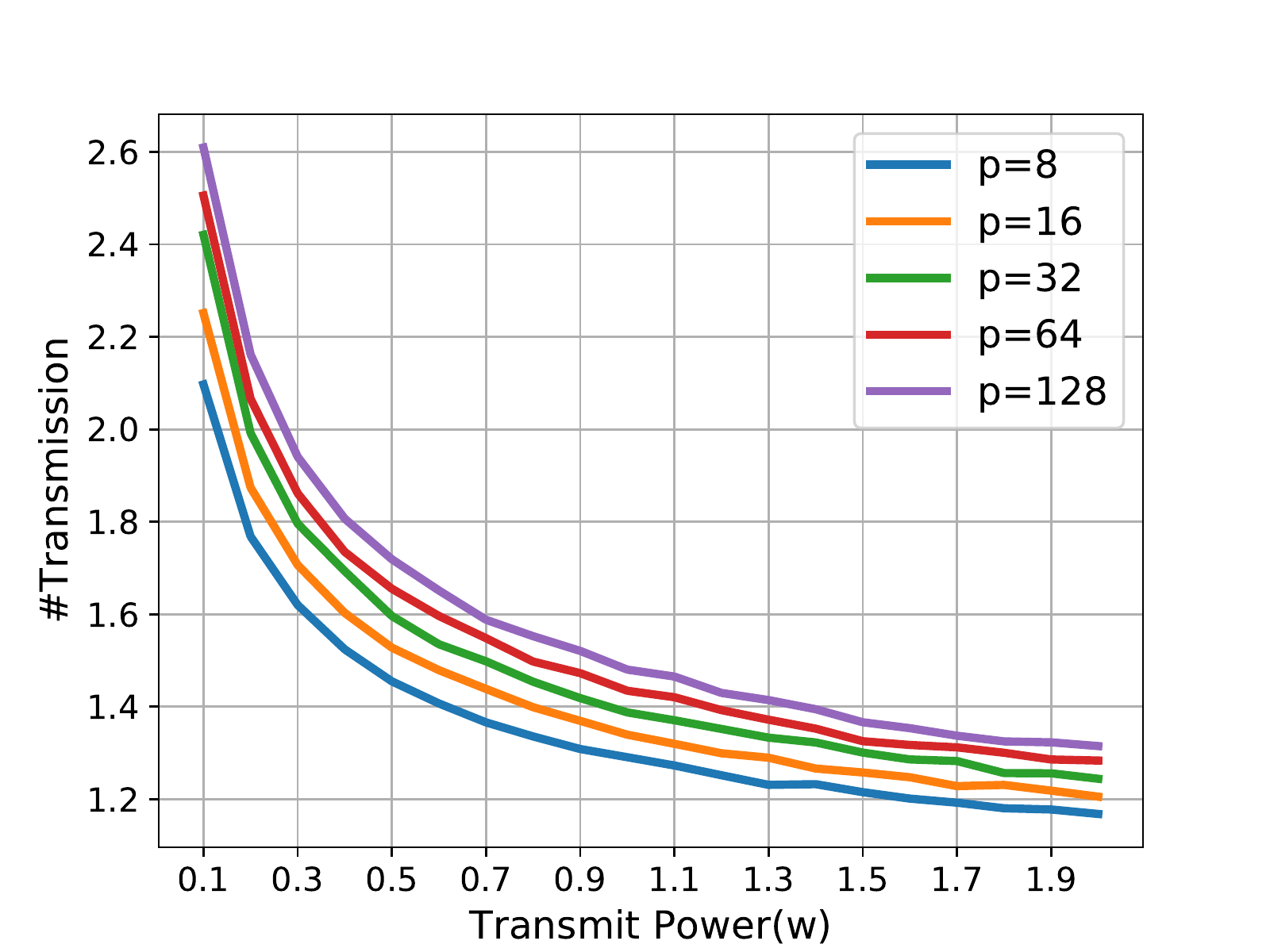}
			\vspace{-1em}
			\label{fig:bit_p_trans}
		\end{minipage}
	}
	\vspace{-0.5em}
	\caption{Simulation results in uncoded communication systems with different node feature dimensions.}
	\label{fig:bit_p}
	\vspace{-2em}
\end{figure}

\begin{figure}
	\vspace{-2em}
	\centering
	\subfigure[\scriptsize{Misclassification rate w/o retransmission.}]{
		\begin{minipage}[t]{0.9\linewidth}
			\centering
			\includegraphics[width=0.8\linewidth]{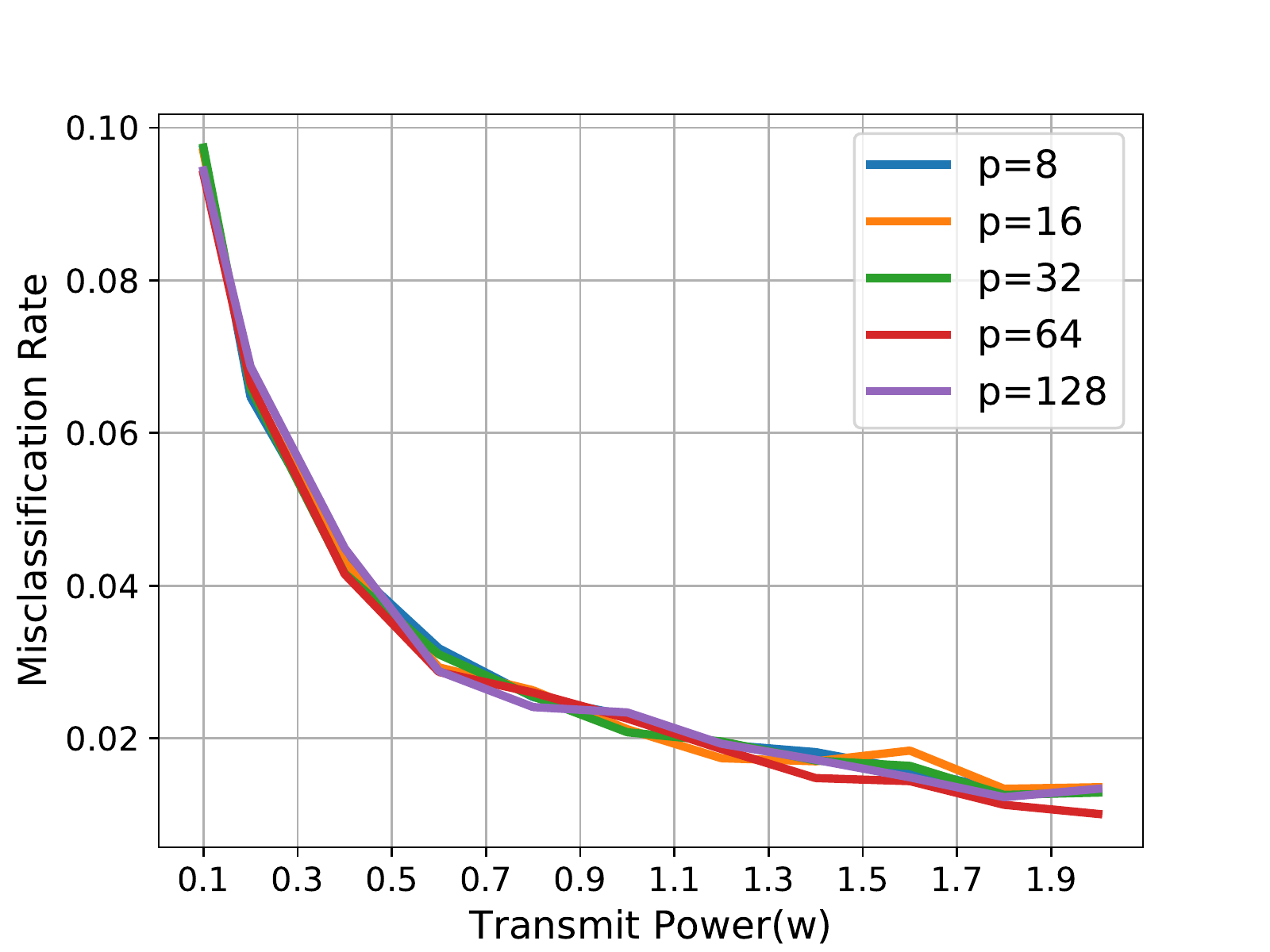}
			\vspace{-1em}
			\label{fig:fading_p_miss}
		\end{minipage}
	}
	\subfigure[\scriptsize{Percentage of robust nodes w/o retransmission.}]{
		\begin{minipage}[t]{0.9\linewidth}
			\centering
			\includegraphics[width=0.8\linewidth]{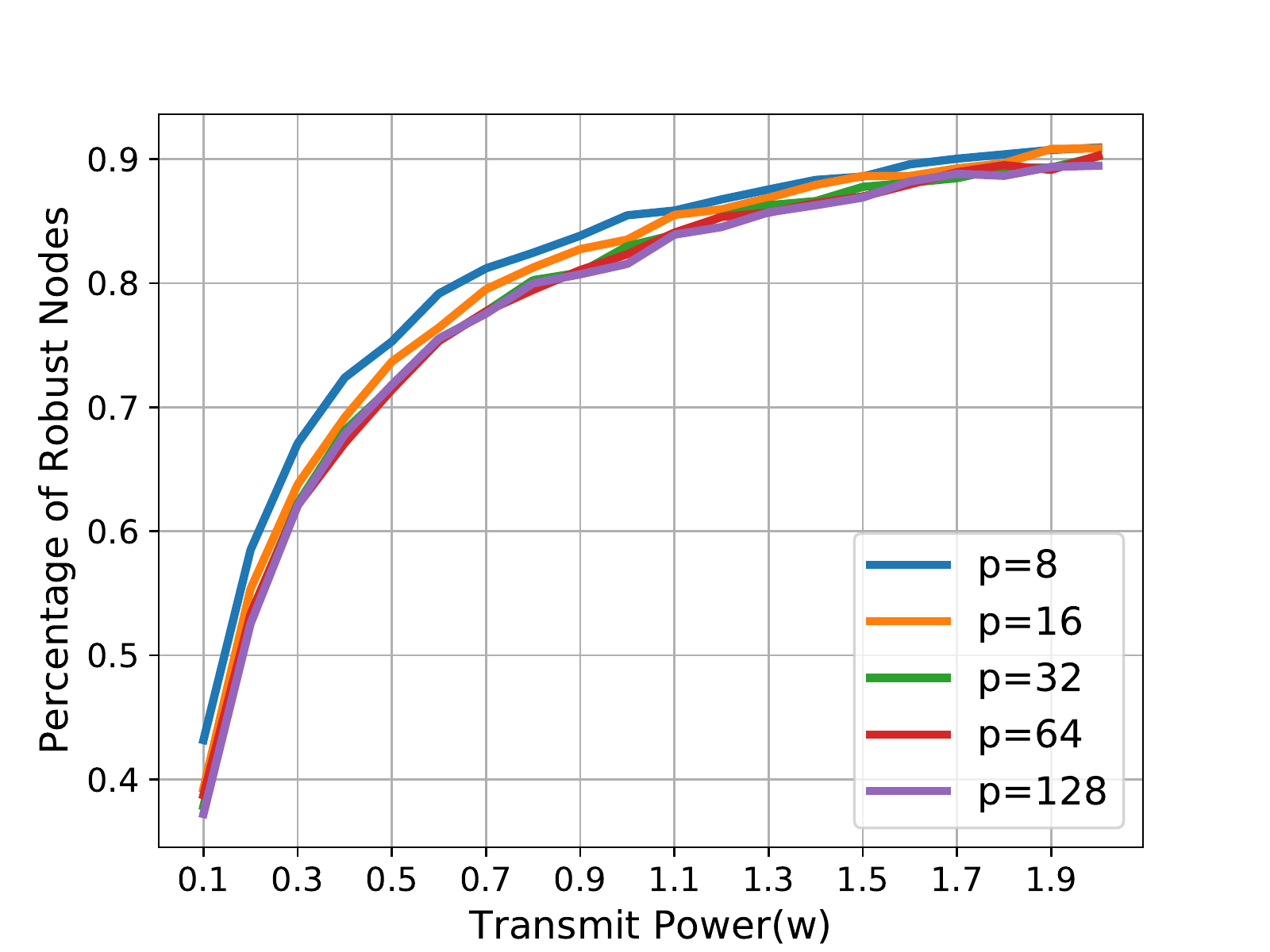}
			\vspace{-1em}
			\label{fig:fading_p_robust}
		\end{minipage}
	}
	\subfigure[\scriptsize{Average needed transmission rounds.}]{
		\begin{minipage}[t]{0.9\linewidth}
			\centering
			\includegraphics[width=0.8\linewidth]{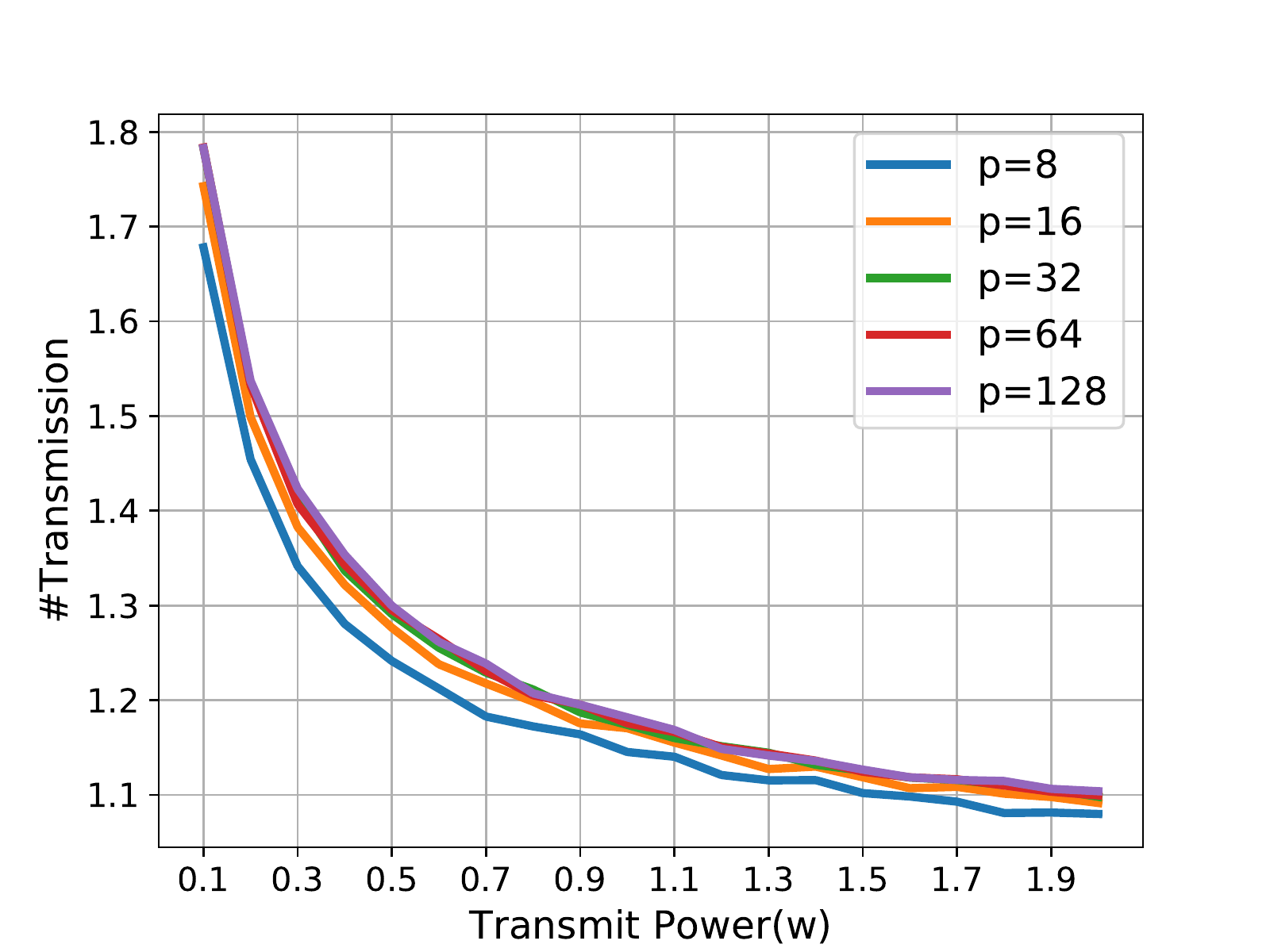}
			\vspace{-1em}
			\label{fig:fading_p_trans}
		\end{minipage}
	}
	\vspace{-0.5em}
	\caption{Simulation results in coded communication systems with different node feature dimensions.}
	\label{fig:fading_p}
	\vspace{-0.5em}
\end{figure}

Besides the node density mentioned above, the node feature dimension is also a key parameter for the practical network design, which directly corresponds to the length of the packet to be transmitted. In the following, we study the impact of the node feature dimension on the decentralized GNN binary classifier in wireless communication systems with $N=50$. The simulation results are summarized in Figs. \ref{fig:bit_p} and \ref{fig:fading_p}.

As can be seen, the impact of the node feature dimension for uncoded and coded communication systems are different. From Fig. \ref{fig:bit_p}, larger node feature dimensions deteriorate the robustness of the decentralized GNN binary classifier in uncoded communication systems (larger misclassification rates and smaller percentages of robust nodes without retransmission). Again, this can be explained based on the definition of robustness probability in (\ref{pr}). From (\ref{pr}), when SINR or BER of the received signal is fixed, increasing $p$ leads to smaller robustness probability. Accordingly, more transmission rounds are needed for networks with larger node feature dimensions to achieve a given robustness requirement, as depicted in Fig. \ref{fig:bit_p_trans}.

However, the robustness of the decentralized GNN binary classifier in coded communication systems does not change with $p$. The reason is that packet-level errors are considered in coded communication systems and the length of each packet does not matter (so long as its transmission can be completed within the coherence time).  Based on the above simulation results, for uncoded communication systems, we need to constraint the node feature dimension. As mentioned in Section \ref{s3_1}, the node feature dimension is  related to the quantization precision.  Higher-precision quantization on the internal characteristics leads to larger node feature dimension but alleviates the difficulty of learning a good GNN model. Therefore, there exists a trade-off between the training difficulty and the prediction robustness for the decentralized GNN binary classifier in uncoded communication systems. 

\subsection{Comparison between Uncoded and Coded Communication Systems}\label{s8_5}
\noindent Up to this point, we have analyzed and tested the performance of the decentralized GNN binary classifier and the corresponding retransmission mechanisms in both uncoded and coded communication systems. As a summary of the discussion in this section, we highlight some major differences between the two communication systems in Table \ref{comparison}. Overall speaking, coded communication systems are superior to uncoded communication systems for the robust implementation of the GNN binary classifier. Undoubtedly, transmission with coded signals induces overhead in other aspects, such as the decoding delay. Choosing specific transmission mode is important for specific wireless applications.
\begin{table}
	\scriptsize
	\caption{Comparison between Uncoded and Coded Communication Systems}
	\vspace{-1em}
	\label{comparison}
	\centering
	\begin{tabular}{|c|c|c|}
		\hline
		System Model& Unoded& Coded\\
		\hline
		Error & Bit-level (BER)&Packet-level (Outage)\\
		\hline
		Robustness Metric&Robustness Probability&Robustness Indicator\\
		\hline
		\tabincell{c}{Misclassification Rate}& Larger & Smaller \\
		\hline
		\tabincell{c}{Percentage of \\Robust Nodes}& Larger & Smaller \\
		\hline
		\tabincell{c}{Average Needed \\Retransmission Rounds}&More & Fewer \\
		\hline
		\tabincell{c}{Scalability on Node\\ Feature Dimension}& Limited & Strong \\
		\hline
	\end{tabular}
	\vspace{-2em}
\end{table}

\section{Conclusion} \label{s9}
\noindent GNN is a potentially powerful tool for a wide range of applications in wireless communication systems. Different from other widely used neural network models, the inference stage of the GNN can be naturally used in a decentralized manner with information exchanges among neighbors over wireless channels. However, wireless transmission is generally imperfect due to the noise and fading impairments, which is the main bottleneck of the decentralized GNN. Therefore, by using the GNN binary classifier as an example, this paper analyzes the performance the decentralized GNN during the inference stage in wireless communication systems and develops retransmission mechanisms to enhance the prediction robustness. We first develop a methodology to check whether the predicted labels are robust. Then we analyze the performance of the decentralized GNN binary classifier in both uncoded and coded communication systems. Furthermore, we develop novel retransmission mechanisms to enhance the prediction robustness in the above two communication systems. To verify our analysis and the effectiveness of the proposed retransmission mechanisms, we conduct simulations on synthetic graph data. The results suggest that our proposed retransmission mechanisms efficiently enhance the robustness of the decentralized GNN. We also provide some insights into related wireless networks design and practical implementation. The proposed analysis, retransmission mechanisms, and corresponding insights are applicable to any GNN based decentralized binary control/management in wireless communication systems.

As far as we know, we are among the first to study the robust decentralized inference with GNNs over noisy wireless channels. The current work represents a first step towards this goal and some open problems still need further investigation. First, this paper focuses on a single-layer GNN binary classifier, where  nodes only need to exchange node features with each other. For multi-layer GNNs, however, hidden states at different layers also need to be shared among neighbors through wireless channels, for which the synchronization among neighbors cannot be neglected. Therefore, designing effective synchronization protocol is an important future direction to extend the analysis in this paper.  Furthermore, we assume the existence of a well-trained GNN model and the accuracy of the decentralized GNN with perfect transmission is 100\% to simplify the analysis. However, this assumption may not hold in practice. Besides imperfect transmission, incorrect predictions may be  caused by  the limited generalizability of the GNN model on unseen samples as well. How to analyze the robustness of the decentralized GNN with limited generalizability to testing samples is worth future investigation. Moreover, all unqualified neighbors are asked to re-transmit in our proposed retransmission mechanism and the robustness is re-checked after all retransmission is done. In practice, the retransmission is generally done in serial. It is possible that after retransmission from some (but not all) unqualified neighbors, the robustness can be achieved and the order of retransmission may matter. Investigating reasonable order of retransmission to reduce the communication overhead is an interesting topic. Finally, we only consider uncoded and simplified coded communication systems in this paper for theoretical analysis. However, error correction codes, cyclic redundancy check or other more complicated coding mechanisms which can help detect and correct errors are usually adopted in practical systems. These mechanisms definitely influence the robustness of the decentralized GNN. One can also consider other techniques, such as the joint source-channel coding and the error correction mechanism, instead of the retransmission mechanism to further enhance the robustness of the decentralized GNN in wireless communication systems.

\appendices
\section{Dual Problem for Problem (23)} \label{appendix_a}
\noindent We reformulate Problem (23) as follows 
\begin{equation}\label{Mprimal}
\min_{\{\bm{\tilde{X}}_v,\bm{h}_v,\bm{\hat{h}_v},\tilde{c}_v,\bm{\Delta}_v\}} \hat{c}_v \times \tilde{c}_v,
\end{equation}
subject to 
\vspace{-1em}
\begin{align}
\tilde{c}_v = \bm{h}_v\bm{w}+b, \Rightarrow \rho_v \in \mathbb{R} \label{primal_1}\tag{\theequation a}
\end{align}
\vspace{-2em}
\begin{align}
\bm{\hat{h}_v} = \hat{a}_v\bm{x}_v\bm{\theta}+\bm{\hat{a}}_v\bm{\tilde{X}}_v\bm{\theta}, \Rightarrow \bm{\alpha}_v \in \mathbb{R}^{1\times D}  \label{primal_2}\tag{\theequation b}
\end{align}
\vspace{-2em}
\begin{align}
0 \leq \bm{\tilde{X}}_v[u,j] \leq 1, \forall u\in N(v),  j \in \{1, 2, .., p\}, \notag \\ \Rightarrow \bm{\Psi}_v^-, \bm{\Psi}_v^+ \in \mathbb{R}^{|N(v)|\times p} \label{primal_3}\tag{\theequation c}
\end{align}
\vspace{-1.5em}
\begin{align}
||\bm{\tilde{X}}_v[u,j]-\bm{\hat{X}}_v[u,j]|| \leq \bm{\Delta}_v[u,j], \forall u\in N(v),  \notag \\ j \in \{1, 2, .., p\}, \Rightarrow \bm{\Upsilon}_v^+, \bm{\Upsilon}_v^- \in \mathbb{R}^{|N(v)|\times p} \label{primal_6}\tag{\theequation d}
\end{align}
\vspace{-2em}
\begin{align}
\sum_{j=1}^p\bm{\Delta}_v[u,j] \leq q_{vu}, \forall u\in N(v), \Rightarrow \bm{\beta}_v \in \mathbb{R}^{|N(v)|}\label{primal_7}\tag{\theequation e}
\end{align}
\vspace{-1.5em}
\begin{align}
\bm{h}_v[i]=\bm{\hat{h}_v}[i], \forall  i \in \mathcal{I}_v^+, \quad
\bm{h}_v[i]=0, \forall  i \in \mathcal{I}_v^-,  \label{primal_9}\tag{\theequation f}
\end{align}
\vspace{-2em}
\begin{align}
\bm{h}_v[i] \geq 0,\forall  i \in \mathcal{I}_v, \Rightarrow \bm{\tau}_v \in \mathbb{R}^{1\times D} \label{primal_10}\tag{\theequation g}
\end{align}
\vspace{-2em}
\begin{align}
\bm{h}_v[i] \geq  \bm{\hat{h}_v}[i], \forall  i \in \mathcal{I}_v, \Rightarrow \bm{\mu}_v \in \mathbb{R}^{1\times D}  \label{primal_11}\tag{\theequation h}
\end{align}
\vspace{-2em}
\begin{align}
\bm{h}_v[i](\bm{u}_v[i]-\bm{l}_v[i]) \leq \bm{u}_v[i](\bm{\hat{h}_v}[i]-\bm{l}_v[i]),\forall  i \in \mathcal{I}_v.\notag \\  \Rightarrow \bm{\lambda}_v \in \mathbb{R}^{1\times D} \label{primal_12}\tag{\theequation i}
\end{align}
Note that constraint (\ref{primal_9}) can be simply eliminated from the optimization. Then the dual problem of Problem (\ref{Mprimal}) is given by \vspace{-1em}
\begin{align}\label{Mdual1}
&\max_{\{\rho_v, \bm{\alpha}_v, \bm{\Psi}_v^+, \bm{\Psi}_v^-,\bm{\Upsilon}_v^+, \bm{\Upsilon}_v^-, \bm{\beta}_v,\bm{\tau}_v,\bm{\mu}_v,\bm{\lambda}_v\}} -\rho_vb\notag\\
&-\sum_{i=1}^D \bm{\alpha}_v[i](\hat{a}_v\bm{x}_v\bm{\theta})[i]-\sum_{u\in N(v)}\sum_{j=1}^p \bm{\Psi}_v^+[u,j]\notag \\
&+\sum_{u\in N(v)}\sum_{j=1}^p\bm{\hat{X}}_v[u,j](\bm{\Upsilon}_v^-[u,j]-\bm{\Upsilon}_v^+[u,j])
\notag\\&-\sum_{u\in N(v)} q_{vu}\bm{\beta}_v[u] + \sum_{i\in \mathcal{I}_v} \bm{\lambda}_v[i]\bm{u}_v[i]\bm{l}_v[i], 
\end{align}
subject to 
\begin{align}
\bm{\beta}_v[u] \geq \bm{\Upsilon}_v^+[u,j]+\bm{\Upsilon}_v^-[u,j], \forall u\in N(v),  j \in \{1, 2, .., p\},\label{dual1_1}\tag{\theequation a}
\end{align}
\vspace{-1.5em}
\begin{align}
\bm{\hat{a}}_v^T \bm{\alpha}_v\bm{\theta}^T=\bm{\Psi}_v^+-\bm{\Psi}_v^-+\bm{\Upsilon}_v^+-\bm{\Upsilon}_v^-,\label{dual1_2}\tag{\theequation b}
\end{align}
\vspace{-2em}
\begin{align}
\rho_v = -\hat{c}_v,\label{dual1_3}\tag{\theequation c}
\end{align}
\vspace{-2em}
\begin{align}
\bm{\alpha}_v[i] = \bm{\lambda}_v[i]\bm{u}_v[i]-\bm{\mu}_v[i], \forall i\in \mathcal{I}_v,\label{dual1_4}\tag{\theequation d}
\end{align}
\vspace{-2em}
\begin{align}
(\rho_v\bm{w}^T)[i]=\bm{\lambda}_v[i](\bm{u}_v[i]-\bm{l}_v[i]) - \bm{\tau}_v[i] -\bm{\mu}_v[i], \forall i\in \mathcal{I}_v\label{dual1_5}\tag{\theequation e}
\end{align}
\vspace{-1.5em}
\begin{align}
\bm{\alpha}_v[i]=0, \forall i\in \mathcal{I}_v^-, \quad
\bm{\alpha}_v[i]=(\rho_v\bm{w}^T)[i], \forall i\in \mathcal{I}_v^+,\label{dual1_7}\tag{\theequation f}
\end{align}
\vspace{-2em}
\begin{align}
\bm{\Psi}_v^+, \bm{\Psi}_v^-,\bm{\Upsilon}_v^+, \bm{\Upsilon}_v^-, \bm{\beta}_v,\bm{\tau}_v,\bm{\mu}_v,\bm{\lambda}_v\succeq 0. \label{dual1_8}\tag{\theequation g}
\end{align}
\vspace{-2em}
		
We know that either $\bm{\lambda}_v$ or $\bm{\tau}_v+\bm{\mu}_v$ must be zero, which correspond to upper and lower bound of $\bm{h}_v$, respectively.  According to (\ref{dual1_5}), if $\bm{\lambda}_v[i]=0$, then we have 	$$(\rho_v\bm{w}^T)[i]=- \bm{\tau}_v[i] -\bm{\mu}_v[i] \Rightarrow  \bm{\tau}_v[i]+\bm{\mu}_v[i]=[(\rho_v\bm{w}^T)[i]]_-. \vspace{-0.5em}$$
Similarly, if $\bm{\tau}_v[i]+\bm{\mu}_v[i]=0$,
$$\bm{\lambda}_v[i](\bm{u}_v[i]-\bm{l}_v[i]) =[(\rho_v\bm{w}^T)[i]]_+. $$
Therefore, we have $\bm{\lambda}_v[i]=[(\rho_v\bm{w}^T)[i]]_+/(\bm{u}_v[i]-\bm{l}_v[i])$. Combined (\ref{dual1_3}), the following term in the objective function of Problem (\ref{Mdual1}) can be rewritten as 
$\bm{\lambda}_v[i]\bm{u}_v[i]\bm{l}_v[i]=\frac{\bm{u}_v[i]\bm{l}_v[i]}{\bm{u}_v[i]-\bm{l}_v[i]}[\hat{c}_v\bm{w}[i]]_-, $ which is constant.
	
To further simplify the objective function of Problem (\ref{Mdual1}), we introduce a new vector, $\bm{\phi}_v \in [0,1]^D$, which satisfies $\bm{\mu}_v[i]=\bm{\phi}_v[i][(\rho_v\bm{w}^T)[i]]_-$. Therefore, constraint (\ref{dual1_4}) can be rewritten as 
\begin{equation}
\begin{aligned}
\bm{\alpha}_v[i] &= \bm{\lambda}_v[i]\bm{u}_v[i]-\bm{\mu}_v[i]\\&=\frac{\bm{u}_v[i]}{\bm{u}_v[i]-\bm{l}_v[i]}[(\rho_v\bm{w}^T)[i]]_+-\bm{\phi}_v[i][(\rho_v\bm{w}^T)[i]]_-. \label{obj2} 
\end{aligned}
\end{equation}
Similarly, we introduce a new matrix, $\bm{\Theta}_v \in [0,1]^{|N(v)|\times p}$, which satisfies the following equations
\begin{equation*}  
\left\{  
\begin{array}{lr}
\bm{\Psi}_v^+[u,j] = \bm{\Theta}_v[u,j][(\bm{\hat{a}}_v^T \bm{\alpha}_v\bm{\theta}^T)[u,j]]_+, &\\
\bm{\Psi}_v^-[u,j] = \bm{\Theta}_v[u,j][(\bm{\hat{a}}_v^T \bm{\alpha}_v\bm{\theta}^T)[u,j]]_-, &\\
\bm{\Upsilon}_v^+[u,j] =(1- \bm{\Theta}_v[u,j])[(\bm{\hat{a}}_v^T \bm{\alpha}_v\bm{\theta}^T)[u,j]]_+, &\\
\bm{\Upsilon}_v^-[u,j] =(1- \bm{\Theta}_v[u,j])[(\bm{\hat{a}}_v^T \bm{\alpha}_v\bm{\theta}^T)[u,j]]_-.&\\
\end{array}
\right.
\end{equation*} 
In this way, constraint (\ref{dual1_1}) can be rewritten as
$\bm{\beta}_v[u] \geq (1- \bm{\Theta}_v[u,j])|(\bm{\hat{a}}_v^T \bm{\alpha}_v\bm{\theta}^T)[u,j]|.$
Based on the above transformation, we know that
$\bm{\Theta}_v[u,j]\geq 1- \bm{\beta}_v[u] /|(\bm{\hat{a}}_v^T \bm{\alpha}_v\bm{\theta}^T)[u,j]|,$
and the following term in the objective function of Problem (\ref{Mdual1}) can be rewritten as \vspace{-0.5em}
\begin{equation}
\begin{aligned}
\bm{\hat{X}}_v[u,j](\bm{\Upsilon}_v^-[u,j]-\bm{\Upsilon}_v^+[u,j])-\bm{\Psi}_v^+[u,j] =\\ -\bm{\hat{X}}_v[u,j](\bm{\hat{a}}_v^T \bm{\alpha}_v\bm{\theta}^T)[u,j]-\bm{\Theta}_v[u,j]\bm{\varepsilon}_v[u,j], \label{tmp1}
\end{aligned}
\end{equation}
where 
\begin{equation}
\begin{aligned} \bm{\varepsilon}_v[u,j]=(1-\bm{\hat{X}}_v[u,j])[(\bm{\hat{a}}_v^T\bm{\alpha}_v\bm{\theta}^T)[u,j]]_+\\+\bm{\hat{X}}_v[u,j][(\bm{\hat{a}}_v^T\bm{\alpha}_v\bm{\theta}^T)[u,j]]_-.
\end{aligned}
\end{equation}
In fact, (\ref{tmp1}) can be transformed into a more tractable form. When $(\bm{\hat{a}}_v^T\bm{\alpha}_v\bm{\theta}^T)[u,j]>0$, if $\bm{\hat{X}}_v[u,j]=1$, $\bm{\varepsilon}_v[u,j]=0$ holds. Similarly, if $\bm{\hat{X}}_v[u,j]=0$,  $\bm{\varepsilon}_v[u,j]=(\bm{\hat{a}}_v^T\bm{\alpha}_v\bm{\theta}^T)[u,j]$. Under this condition, to maximize the objective function, $\bm{\Theta}_v[u,j]$ should be set as its lower bound and \vspace{-0.5em}
\begin{equation}
\begin{aligned}
\bm{\varepsilon}_v[u,j]\bm{\Theta}_v[u,j]&=(\bm{\hat{a}}_v^T\bm{\alpha}_v\bm{\theta}^T)[u,j]\{1- \frac{\bm{\beta}_v[u] }{|(\bm{\hat{a}}_v^T \bm{\alpha}_v\bm{\theta}^T)[u,j]|}\}\\&=\bm{\varepsilon}_v[u,j]-\bm{\beta}_v[u]. 
\end{aligned}
\end{equation} 
Similar analysis can be conducted when $(\bm{\hat{a}}_v^T\bm{\alpha}_v\bm{\theta}^T)[u,j]<0$. Overall, we have 
$\bm{\varepsilon}_v[u,j]\bm{\Theta}_v[u,j]=
\max\{\bm{\varepsilon}_v[u,j]-\bm{\beta}_v[u], 0\}\triangleq \bm{\Omega}_v[u,j].$
Therefore, (\ref{tmp1}) can be rewritten as
\begin{equation*}
\begin{aligned}
&\bm{\hat{X}}_v[u,j](\bm{\Upsilon}_v^-[u,j]-\bm{\Upsilon}_v^+[u,j])-\bm{\Psi}_v^+[u,j]\\ &= -\bm{\hat{X}}_v[u,j](\bm{\hat{a}}_v^T \bm{\alpha}_v\bm{\theta}^T)[u,j]-\max\{\bm{\varepsilon}_v[u,j]-\bm{\beta}_v[u], 0\},\\
&= -\text{tr}[\bm{\hat{X}}_v^T(\bm{\hat{a}}_v^T\bm{\alpha}_v\bm{\theta}^T)]-||\bm{\Omega}_v||_1,
\end{aligned} 
\end{equation*}
where $\text{tr}(\cdot)$ indicates the trace of the input matrix. Based on the above transformations, Problem (\ref{Mdual1}) finally can be transformed into Problem (24) in Section 4.3.
	
\section{Solution for Problem (26)} \label{appendix_b}
\noindent Problem (26) can be reformulated as \vspace{-0.5em}
\begin{equation}\label{Mduals1}
\min_{\{\bm{\beta}_v\succeq 0,\bm{\hat{\Omega}}_v\succeq 0\}} \sum_{u\in N(v)}\sum_{j=1}^p \bm{\hat{\Omega}}_v[u,j]+\sum_{u\in N(v)}q_{vu}\bm{\beta}_v[u]-\text{const}, 
\end{equation}
subject to 
\begin{align}
\bm{\hat{\Omega}}_v[u,j] \geq \bm{\varepsilon}_v[u,j]-\bm{\beta}_v[u], \forall u\in N(v),  j \in \{1, 2, .., p\}, \label{duals1_2}\tag{\theequation a}
\end{align}
\vspace{-2em}
\begin{align}
\text{const} = &c_vb+ \sum_{i\in \mathcal{I}_v}\frac{\bm{u}_v[i]\bm{l}_v[i]}{\bm{u}_v[i]-\bm{l}_v[i]}[c_v\bm{w}[i]]_-\notag\\&-\sum_{i=1}^D\bm{\alpha}_v[i](\hat{a}_v\bm{x}_v\bm{\theta})[i]-\text{tr}[\bm{X}_v^T(\bm{\hat{a}}_v^T\bm{\alpha}_v\bm{\theta}^T)]. \label{duals1_3}\tag{\theequation b}
\end{align}
The dual problem of Problem (\ref{Mduals1}) is given by \vspace{-0.5em}
\begin{equation}\label{Mduals2}	
\max_{\bm{B}_v} \sum_{u\in N(v)}\sum_{j=1}^p \bm{\varepsilon}_v[u,j]\bm{B}_v[u,j]-\text{const}, 
\end{equation}
subject to (\ref{duals1_3}),
\begin{align}
0 \leq \bm{B}_v[u,j] \leq 1, \forall u\in N(v),  j \in \{1, 2, .., p\}, \label{duals2_1}\tag{\theequation a}
\end{align}
\vspace{-2em}
\begin{align}
\sum_{j=1}^p \bm{B}_v[u,j] \leq q_{vu}, \forall u\in N(v). \label{duals2_2}\tag{\theequation b}
\end{align}
The optimal solution of the above problem is very obvious. Specifically, for $\bm{B}_v[u,:]$, its optimal solution is obtained by setting $\bm{B}_v[u,j]$ corresponding to the top-$q_{vu}$ largest entries in $\bm{\varepsilon}_v[u,:]$ as 1 and others as 0.
	
Because $\bm{B}_v$ is the dual variable corresponding to constraint (\ref{duals1_2}), the following equation holds
\begin{equation}
\begin{aligned}
\bm{B}_v[u,j](\bm{\varepsilon}_v[u,j]-\bm{\beta}_v[u]-\bm{\hat{\Omega}}_v[u,j])=0,\\ \forall u\in N(v),  j \in \{1, 2, .., p\}. \label{tmp3}
\end{aligned}
\end{equation}
If $\bm{B}_v[u,j]=1$, $\bm{\hat{\Omega}}_v[u,j]=\bm{\varepsilon}_v[u,j]-\bm{\beta}_v[u]$ holds, which indicates that $\bm{\varepsilon}_v[u,j]-\bm{\beta}_v[u] \geq 0$. Meanwhile, $\bm{B}_v[u,j]=1$ also means that $\bm{\varepsilon}_v[u,j]$ is among the top-$q_{vu}$ largest entries in $\bm{\varepsilon}_v[u,:]$. Therefore, $\bm{\beta}_v[u]$ is no larger than all top-$q_{vu}$ largest entries in $\bm{\varepsilon}_v[u,:]$. Similarly, if $\bm{\varepsilon}_v[u,j]-\bm{\beta}_v[u]-\bm{\hat{\Omega}}_v[u,j] \neq 0$, it must be the case that $\bm{\hat{\Omega}}_v[u,j]=0$ and $\bm{\varepsilon}_v[u,j]-\bm{\beta}_v[u]<0$. According to (\ref{tmp3}), $\bm{B}_v[u,j]$ must equal to 0, which suggests that $\bm{\varepsilon}_v[u,j]$  is not among the top-$q_{vu}$ largest entries in $\bm{\varepsilon}_v[u,:]$ under this condition. Therefore, $\bm{\beta}_v[u]$ should be larger than all the $\bm{\varepsilon}_v[u,j]$  that is not among the top-$q_{vu}$ largest entries in $\bm{\varepsilon}_v[u,:]$. Overall, the solution of $\bm{\beta}_v[u]$ is the $q_{vu}$-largest entry in $\bm{\varepsilon}_v[u,:]$.

\end{document}